\documentclass[journal]{IEEEtran}

\usepackage{graphicx,psfrag,epsfig,epsf,latexsym,hhline,amsmath,amssymb,multirow}
\usepackage{pst-plot}
\usepackage{pstricks-add}
\usepackage{pifont}
\usepackage{graphicx}
\usepackage{amsthm}
\usepackage[linesnumbered,ruled,vlined]{algorithm2e}
\usepackage[noadjust]{cite}
\usepackage{array}
\usepackage{blindtext}
\usepackage{etoolbox}
\usepackage{comment}
\usepackage{epstopdf}
\usepackage{hyperref}

\newcolumntype{M}[1]{>{\centering\arraybackslash}m{#1}}
\newcolumntype{P}[1]{>{\centering\arraybackslash}p{#1}}

\newcommand{\iid}{i.\@i.\@d.\ }

\SetKwInput{KwInput}{Input}                
\SetKwInput{KwOutput}{Output}              

\theoremstyle{definition}
\theoremstyle{definition}
\theoremstyle{definition}
\theoremstyle{definition}

\newtheorem{example}{Example}
\newtheorem{remark}{Remark}

\begin{document}
\title{Analysis and Design of Partially Information- and Partially Parity-Coupled Turbo Codes}
\author{Min Qiu,~\IEEEmembership{Member,~IEEE,}
 Xiaowei Wu,~\IEEEmembership{Student Member,~IEEE,}
 Alexandre Graell i Amat,~\IEEEmembership{Senior Member,~IEEE,}
 and Jinhong Yuan~\IEEEmembership{Fellow,~IEEE}


\thanks{This work was presented in part at the 2019 IEEE Information Theory Workshop (ITW), Visby, Gotland, Sweden \cite{8989359}.

M. Qiu, X. Wu and J. Yuan are with the School of Electrical Engineering and Telecommunications, University of New South Wales, Sydney, NSW, 2052 Australia (e-mail: min.qiu@unsw.edu.au; xiaowei.wu@unsw.edu.au; j.yuan@unsw.edu.au).
A. Graell i Amat is with the Department of Electrical Engineering, Chalmers University of Technology, SE-41296 Gothenbury, Sweden (e-mail: alexandre.graell@chalmers.se).
}%
}

\maketitle

\begin{abstract}
In this paper, we study a class of spatially coupled turbo codes, namely partially information- and partially parity-coupled turbo codes. This class of codes enjoy several advantages such as flexible code rate adjustment by varying the coupling ratio and the encoding and decoding architectures of the underlying component codes can remain unchanged. For this work, we first provide the construction methods for partially coupled turbo codes with coupling memory $m$ and study the corresponding graph models. We then derive the density evolution equations for the corresponding ensembles on the binary erasure channel to precisely compute their iterative decoding thresholds. Rate-compatible designs and their decoding thresholds are also provided, where the coupling and puncturing ratios are jointly optimized to achieve the largest decoding threshold for a given target code rate. Our results show that for a wide range of code rates, the proposed codes attain close-to-capacity performance and the decoding performance improves with increasing the coupling memory. In particular, the proposed partially parity-coupled turbo codes have thresholds within 0.0002 of the BEC capacity for rates ranging from $1/3$ to $9/10$, yielding an attractive way for constructing rate-compatible capacity-approaching channel codes.
\end{abstract}


\section{Introduction}
Spatially coupled codes, originally introduced in \cite{LDPCCthesis,782171} as convolutional low-density parity-check (LDPC) codes (also known as spatially coupled LDPC (SC-LDPC) codes), have now been recognized as promising candidates for a range of applications such as optical communication \cite{7340116} and data storage systems \cite{7553579}. SC-LDPC codes can be obtained by spreading the edges of the Tanner graph \cite{1056404} of the underlying uncoupled regular LDPC block codes \cite{Gallager63low-densityparity-check} to several adjacent blocks \cite{5571910}. It was shown in \cite{5571910,5695130,7130575,7152893,7533500} that SC-LDPC codes have much better decoding thresholds than those of the uncoupled LDPC codes. In particular, SC-LDPC codes exhibit a phenomenon called threshold saturation: the decoding threshold under iterative belief propagation (BP) decoding converges to that under maximum-a-posteriori (MAP) decoding for the binary erasure channel (BEC) \cite{5695130,6325197} and the class of binary memoryless symmetric channels in general \cite{6589171,6912949}. In addition, SC-LDPC codes can be efficiently decoded by a sliding window decoder in a component-wise manner \cite{6374679}, enabling a continuous streaming fashion with a lower decoding delay than conventional LDPC block codes. Due to these advantages, the research on SC-LDPC codes has attracted considerable interest in both academia and industry \cite{7265214}.

Spatial coupling techniques have also been applied to other codes. For example, a class of spatially coupled product codes called staircase codes \cite{Smith12,8425763} have demonstrated superior error performance over uncoupled product codes and they have close-to-capacity performance under iterative bounded-distance decoding. Recently, spatially coupled turbo-like codes were introduced in \cite{8002601,8631116}, where the turbo-like codes are referring to the codes whose graph representations \cite{910572} have trellis constraints. Specifically, the component codes used for spatial coupling include parallel concatenated convolutional codes (PCCs) \cite{397441,Vucetic:2000:TCP:352869}, serially concatenated convolutional codes (SCCs) \cite{669119} and braided convolutional codes (BCCs) \cite{5361461}. It was proved that threshold saturation also occurs for all the proposed spatially coupled turbo-like codes in \cite{8002601}. However, it can still be observed that spatially coupled PCCs (SC-PCCs) have a noticeable gap to the capacity for various code rates unlike spatially coupled SCCs (SC-SCCs) and spatially coupled BCCs (SC-BCCs). This is because the MAP thresholds of PCCs \cite{8002601} (when punctured) are away from capacity and thus the same happen to the BP thresholds of the resulting SC-PCCs \cite{8002601}. Motivated by the fact that PCCs (also called turbo codes) are the standard channel coding schemes in 3G and 4G wireless mobile communication systems which coexist with 5G systems, we are particular interested in designing and investigating new spatial coupling techniques to further enhance the performance of turbo codes.

Very recently, a class of spatially coupled turbo codes called partially information-coupled turbo codes (PIC-TCs) were proposed in \cite{8368318} to enhance the performance of the hybrid automatic repeat request protocol of 4G \cite{4907407} and 5G \cite{6824752} by reducing the error rate of each transport block (TB) and the number of retransmissions. Instead of using a very long code to encode the entire information of a TB into a codeword, in PIC-TCs the information sequence of a TB is divided into several small sub-sequences and each sub-sequence as well as a part of the information bits from consecutive sub-sequences are encoded into a component codeword. In other words, some of the information bits are shared between consecutive component codewords. This introduced \emph{coupling} between component codewords improves the reliability of the transmitted TBs while the spatial coupling nature of PIC-TCs allows low latency decoding via sliding window decoding. Furthermore, the encoding and decoding of the component codewords is performed by standard turbo encoding and decoding, respectively. It is also worth pointing out that the component code can be any linear code, e.g., LDPC codes \cite{8301547} and polar codes \cite{8470926}, as long as the component code is \emph{systematic}. Due to the partially information coupling on the turbo code level, the code rates of PIC-TCs can be adjusted flexibly by varying the coupling ratio (i.e., the ratio of the number of shared information bits to the number of information bits of a component codeword). Simulation results show that PIC-TCs constructed from LTE turbo codes can have a signal-to-noise ratio gain up to 0.73 dB over uncoupled LTE turbo codes in the Gaussian channel \cite{8368318}. However, only codes with coupling memory $m=1$ and rates lower than $1/3$ were considered in \cite{8368318}. Moreover, in \cite{8368318} an upper bound on the decoding threshold of PIC-TCs was derived via extrinsic information transfer (EXIT) functions \cite{957394} by assuming that the coupled information bits for \emph{every} component code are perfectly known to the decoder. The exact decoding threshold of PIC-TCs, however, was not derived \cite{8368318}. Thus, the theoretical performance of the codes has not been fully understood.

In this paper, we study the design and analyze partially coupled turbo codes. Apart from PIC-TCs, we also propose a new family of partially coupled codes called partially parity-coupled turbo codes (PPC-TCs). The decoding performance of both codes under the BEC are investigated. Our results show that applying the technique of partial coupling to turbo codes can lead to significant threshold improvements over uncoupled turbo codes. Moreover, both PIC-TCs and PPC-TCs are fully compatible with the current wireless systems as we keep the encoding and decoding architectures of the component turbo codes unchanged. The main contributions of the papers are as follows:
\begin{itemize}
\item We first introduce a general construction method of PIC-TCs with coupling memory $m\geq 1$. Inspired by the design of PIC-TCs, we also propose PPC-TCs, a new family of partially coupled codes which extend the concept of PIC-TCs to the coupling of parity sequences and provide a great flexibility in terms of code rates. The decoding procedures and the update rules for the log-likelihood ratio (LLR) for both codes are also provided.

\item We derive the exact density evolution (DE) equations for PIC-TC and PPC-TC ensembles and compute their iterative decoding thresholds on the BEC. For both PIC-TCs and PPC-TCs, we observe that the decoding thresholds improve as the coupling ratio and the coupling memory increase. In particular, PIC-TCs have a gap to the BEC capacity of 0.0018 when half of the information bits of the component codewords are coupled.

\item We investigate the rate-compatible design of PIC-TCs and PPC-TCs and derive the respective DE equations assuming random puncturing. We jointly optimize the coupling and puncturing ratios in order to achieve the largest decoding threshold for a given code rate and coupling memory. We show that by choosing the optimal pair of coupling ratio and puncturing ratio, the proposed PIC-TCs and PPC-TCs achieve larger decoding thresholds than SC-PCCs in \cite{8002601} while PPC-TCs even have better performance than SC-SCCs \cite{8002601}  and comparable performance to SC-BCCs \cite{8002601} with large coupling memory. Most notably, PPC-TCs only show a gap to the BEC capacity within 0.0002 for a wide range of code rates for large enough coupling memory. The design of PPC-TCs can be easily extended to the cases with very low rates (e.g., rate-$1/10$ and lower) by increasing the coupling ratio. Our analysis suggests that it is possible to approach the capacity over a wide range of code rates by applying the partial coupling technique on existing turbo codes without changing their encoding and decoding architectures.

\item We evaluate the error performance of PIC-TCs and PPC-TCs with simulations. The theoretical analysis is verified via simulating the error performance of the proposed PIC-TCs and PPC-TCs. In addition, we also investigate the performance of both codes under sliding window decoding, where we show that window size 4 suffices to allow the proposed codes to achieve considerable error performance and outperform uncoupled turbo codes. We finally show that the optimized PIC-TCs and PPC-TCs also perform well over the AWGN channel with practical component codeword length.
\end{itemize}

\subsection{Notations}
Scalars and vectors are written in lightface and boldface letters, respectively, e.g., $x$ and $\mathbf{x}$. For a vector $\mathbf{x}$,  $n(\mathbf{x})$ gives the number of elements in $\mathbf{x}$.

\section{Partially Information-Coupled Turbo Codes}\label{sec:intro_PIC}
In this section, we introduce the construction of the proposed PIC-TCs. The encoding and decoding procedures of the codes are provided. For ease of understanding, we start with introducing the codes for coupling memory $m=1$ and then show the coupling for higher coupling memories.

In PIC-TCs, linear binary systematic turbo codes are used as the component codes. The partial information coupling means that some information bits are shared between consecutive turbo encoders. The details of the scheme are given in the following.

\begin{figure}[t!]
	\centering
\includegraphics[width=3.42in,clip,keepaspectratio]{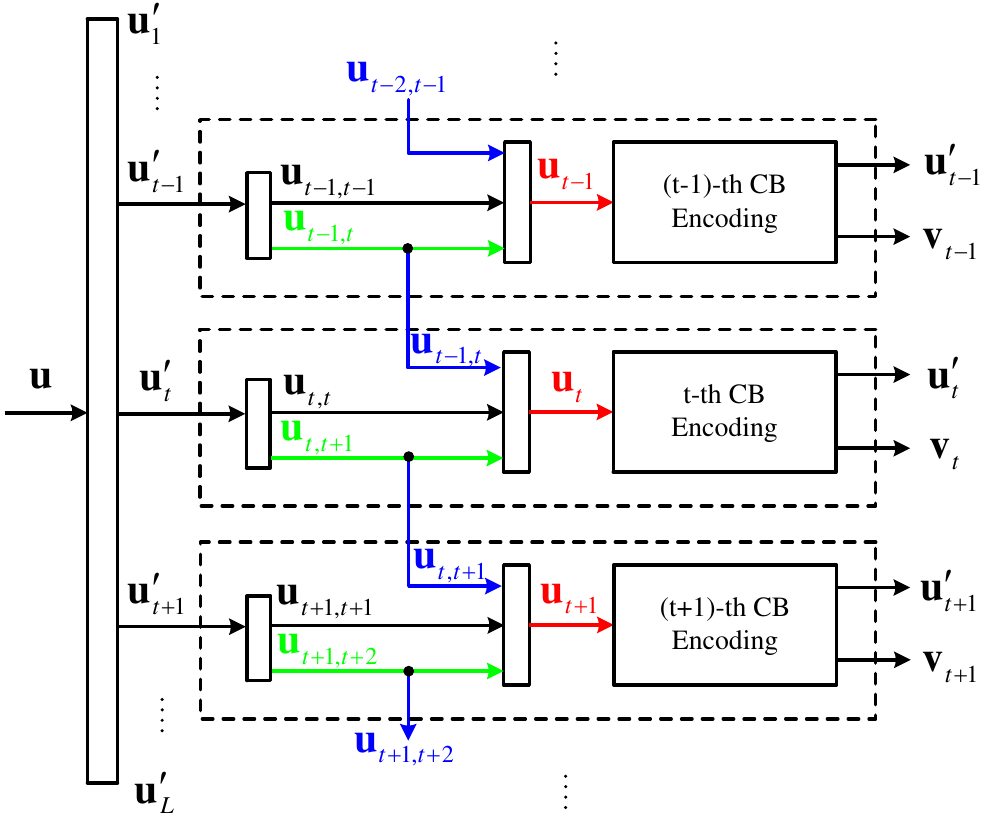}
\caption{Block diagram of the encoder of a PIC-TC with $m=1$.}
\label{fig:1}
\end{figure}

\subsection{Encoding of PIC-TCs}
The block diagram of a PIC-TC with $m = 1$ is depicted in Fig. \ref{fig:1}. The underlying component code is a turbo code with information length $K$ and codeword length $N$. As shown in Fig. \ref{fig:1}, an information sequence $\mathbf{u}$ is divided into $L$ vectors $\mathbf{u}'_1,\ldots,\mathbf{u}'_L$, which are encoded into $L$ code blocks (CBs). Each vector $\mathbf{u}'_t$ for $t\in\{1,\ldots,L\}$ is further decomposed into two vectors $\mathbf{u}_{t,t}$ and $\mathbf{u}_{t,t+1}$, corresponding to the uncoupled and coupled information sequence, respectively. That is, $\mathbf{u}'_t = [\mathbf{u}_{t,t},\mathbf{u}_{t,t+1}]$ as shown in Fig. \ref{fig:1}. Here, $\mathbf{u}_{t,t+1}$ is the coupled information sequence which is a part of the input to both the $t$-th CB encoder and the $(t+1)$-th CB encoder while $\mathbf{u}_{t,t}$ is the uncoupled information sequence that is only encoded by the $t$-th CB encoder. In other words, the information sequence $\mathbf{u}_{t,t+1}$ is shared between the $t$-th CB and the $(t+1)$-th CB and thus this coupled information sequences are encoded \emph{twice} and are protected by two component turbo codewords. For the $t$-th CB encoder, its input is a length $K$ vector $\mathbf{u}_t = [\mathbf{u}_{t-1,t},\mathbf{u}'_{t}] = [\mathbf{u}_{t-1,t},\mathbf{u}_{t,t},\mathbf{u}_{t,t+1}]$. Its output is a length $N$ vector $[\mathbf{u}_t,\mathbf{v}_t] = [\mathbf{u}_{t-1,t},\mathbf{u}'_{t},\mathbf{v}_t]$, where $\mathbf{v}_t$ represents the parity sequence. Considering that the component code is a systematic code, the coupled information $\mathbf{u}_{t-1,t}$, which is encoded twice, is only transmitted once. That is, the final codeword of the $t$-th CB is obtained as $\mathbf{c}_t = [\mathbf{u}'_t,\mathbf{v}_t] = [\mathbf{u}_{t,t},\mathbf{u}_{t,t+1},\mathbf{v}_t]$, where $\mathbf{u}_{t-1,t}$ is punctured at the $t$-th CB since it has already been included as a part of the codeword for the $(t-1)$-th CB, i.e., $\mathbf{c}_{t-1} = [\mathbf{u}_{t-1,t-1},\mathbf{u}_{t-1,t},\mathbf{v}_{t-1}]$.

Recall that the underlying component code is a rate $R_0 =\frac{K}{N}$ turbo code. We denote by $D$ the length of the coupling information sequence $\mathbf{u}_{t,t+1}$. We define the coupling ratio as $\lambda = \frac{D}{K} \in [0,\frac{1}{2}]$, where $\lambda =0$ means that the code is an uncoupled turbo code and $\lambda =\frac{1}{2}$ means that half of the information bits of $\mathbf{u}_t$ are coupled and hence there is no uncoupled information in the codeword, i.e., $n(\mathbf{u}_{t,t}) = 0$. As a result, we have that $n(\mathbf{u}_{t,t+1})=\lambda K,n(\mathbf{u}_{t,t})=(1-2\lambda) K,n(\mathbf{u}'_t)=(1-\lambda) K$, and $n(\mathbf{c}_t)=N-\lambda K$. The coupling ratio is an important parameter that determines the overall code rate and the decoding threshold, which will be discussed later. For initialization and termination of PIC-TCs, zero padding is applied before encoding to the first CB and the $L$-th (last) CB such that $\mathbf{u}_{0,1} = \mathbf{0}$ and $\mathbf{u}_{L,L+1} = \mathbf{0}$, respectively. Then, the code rate of a PIC-TC with $m=1$ is
\begin{align}\label{eq:PICrate_m1}
R_{\text{PIC}} &= \frac{\sum\limits_{t=1}^Ln(\mathbf{u}_t')-n(\mathbf{u}_{L,L+1})}{\sum\limits_{t=1}^Ln(\mathbf{c}_t)- n(\mathbf{u}_{L,L+1})} 
= \frac{L(K-\lambda K)-\lambda K}{L(N-\lambda K)- \lambda K}.
\end{align}

We now consider the general case of coupling memory $m > 1$. At time $t$, the information sequence is decomposed into $\mathbf{u}'_t = [\mathbf{u}_{t,t}, \mathbf{u}_{t,t+1},\ldots,\mathbf{u}_{t,t+m}]$, where for $j \in \{1,\ldots,m\}$, $\mathbf{u}_{t,t+j}$ is the coupled information sequence encoded by both the $t$-th CB encoder and the $(t+j)$-th CB encoder, and $n(\mathbf{u}_{t,t+j}) = \frac{\lambda K}{m}$. The input sequence of the $t$-th CB encoder is the length-$K$ vector $\mathbf{u}_t = [\mathbf{u}_{t-m,t},\ldots,\mathbf{u}_{t-1,t},\mathbf{u}'_{t}] = [\mathbf{u}_{t-m,t},\ldots,\mathbf{u}_{t-1,t},\mathbf{u}_{t,t},\mathbf{u}_{t,t+1},\ldots,\mathbf{u}_{t,t+m}]$. After the turbo encoding, we obtain the length-$N$ component codeword $\mathbf{c}_t = [\mathbf{u}_t',\mathbf{v}_t]=[\mathbf{u}_{t,t},\mathbf{u}_{t,t+1},\ldots,\mathbf{u}_{t,t+m},\mathbf{v}_t]$, where we note that $\mathbf{u}_{t-m,t},\ldots,\mathbf{u}_{t-1,t}$ are punctured at the $t$-th CB. For initialization, zero padding is applied to the coupled information sequences such that $\mathbf{u}_{t-j,t} = \mathbf{0}$ for all $t\in \{1,\ldots,m\}, j \in\{ t,\ldots,m\}$. To terminate PIC-TCs, zero padding is applied to the coupled information sequences such that
$\mathbf{u}_{t,t+j} = \mathbf{0}$ for all $t\in \{L-m+1,\ldots,L\}, j \in\{ L-t+1,\ldots,m\}$. As a result, the code rate of the PIC-TC with coupling memory $m$ is
\begin{align}\label{eq:PICrate2}
R_{\text{PIC}} &= \frac{\sum\limits_{t=1}^Ln(\mathbf{u}_t')-\sum\limits_{t = L-m+1}^L \sum\limits_{j = L-t+1}^m n(\mathbf{u}_{t,t+j})}{\sum\limits_{t=1}^Ln(\mathbf{c}_t)- \sum\limits_{t = L-m+1}^L \sum\limits_{j = L-t+1}^mn(\mathbf{u}_{t,t+j})} \nonumber \\
&= \frac{L(K-\lambda K)-\frac{\lambda K(m+1)}{2}}{L(N-\lambda K)- \frac{\lambda K(m+1)}{2}}.
\end{align}
When $L$ is very large, the code rate in \eqref{eq:PICrate2} becomes
\begin{align}\label{eq:PICrate3}
\lim_{L \rightarrow \infty}R_{\text{PIC}}= \frac{K-\lambda K}{N-\lambda K}
= R_0\frac{(1-\lambda)}{1-\lambda R_0},
\end{align}
where we recall that $R_0 = \frac{K}{N}$ is the code rate of the underlying turbo code.

\begin{remark}
It can be seen from \eqref{eq:PICrate2} that the code rate of a PIC-TC can be lowered by increasing the coupling ratio $\lambda$. Moreover, the structural difference between PIC-TCs and SC-PCCs \cite{8002601} is that the coupling of PIC-TCs is on the turbo code level while for SC-PCCs is on the convolutional code level. This is because the coupled information bits of PIC-TCs are encoded by two turbo encoders (i.e., four convolutional code encoders) while the coupled information bits for SC-PCCs \cite{8002601} are encoded by two convolutional code encoders. As a result, the information nodes in the code graphs of PIC-TCs have \emph{irregular} degrees. Like the case for irregular LDPC codes \cite{910578}, the irregularity in the degree distributions greatly improves the decoding thresholds of the proposed codes as we will see in the density evolution results. Due to the nature of the proposed partial coupling, any linear systematic code can be used as the component code \emph{without changing} its encoder and decoder. It is also worth pointing out that parallel encoding (i.e., encoding all CBs at the same time after coupling) can be used for PIC-TCs to reduce the encoding latency.
\end{remark}

\subsection{Decoding of PIC-TCs}\label{sec:dec_PIC}
\begin{figure*}[t!]
	\centering
\includegraphics[width=6.5in,clip,keepaspectratio]{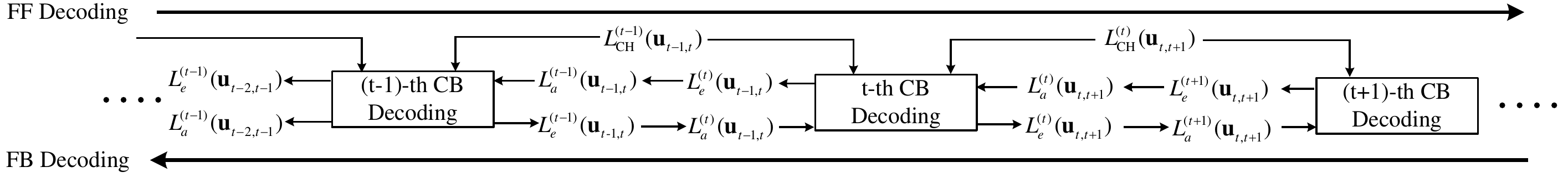}
\caption{An illustration of FF-FB decoding.}
\label{fig:dec_1}
\end{figure*}

The decoding of PIC-TCs is accomplished by a feed-forward and feedback (FF-FB) decoding \cite{8368318} in an iterative manner. Specifically, the FF-FB decoding employs a serial scheduling by decoding the 1st CB to the $L$-th CB serially during the FF decoding and then starts from decoding the $L$-th CB to the 1st CB during the FB decoding, as shown in Fig. \ref{fig:dec_1}. The decoding for each CB uses the component code decoder. In our case, the CB decoding is realized by a standard turbo decoder with the Bahl–Cocke–Jelinek–Raviv (BCJR) decoder \cite{1055186} as the constituent decoder.

The input to the turbo decoder a sequence of $N$ LLRs. For $t \in \{1,\ldots,L\}$, we use $L^{(t)}_{a}(\mathbf{u})$ and $L^{(t)}_{e}(\mathbf{u})$ to represent the \emph{a priori} and extrinsic LLR sequences associated with sequence $\mathbf{u}$ at the $t$-th CB decoder, respectively. We let $L_{\text{CH}}^{(t)}(\mathbf{u})$ to represent the channel LLR sequence associated with sequence $\mathbf{u}$ at the $t$-th CB decoder. To avoid repetition, we focus on the input LLR updates for the \emph{coupled} information bits since the input LLR updates for the uncoupled information bits $\mathbf{u}_{t,t}$ and the parity bits $\mathbf{v}_{t}$ are the same as that for uncoupled turbo codes. The LLRs of the coupled information bits input to the $t$-th CB decoder are updated as follows:
\begin{align}
L^{(t)}&(\mathbf{u}_{t-j,t}) = L_e^{(t)}(\mathbf{u}_{t-j,t})+ L_a^{(t)}(\mathbf{u}_{t-j,t})+ L_{\text{CH}}^{(t-j)}(\mathbf{u}_{t-j,t}) \nonumber \\
&= L_e^{(t)}(\mathbf{u}_{t-j,t})+ L_e^{(t-j)}(\mathbf{u}_{t-j,t})+ L_{\text{CH}}^{(t-j)}(\mathbf{u}_{t-j,t}),\label{eq:LLR1}\\
L^{(t)}&(\mathbf{u}_{t,t+j}) = L_{e}^{(t)}(\mathbf{u}_{t,t+j})+ L_{a}^{(t)}(\mathbf{u}_{t,t+j})+L_{\text{CH}}^{(t)}(\mathbf{u}_{t,t+j})\nonumber \\
&= L_{e}^{(t)}(\mathbf{u}_{t,t+j})+ L_{e}^{(t+j)}(\mathbf{u}_{t,t+j})+L_{\text{CH}}^{(t)}(\mathbf{u}_{t,t+j}).\label{eq:LLR2}
\end{align}
where we note that for $j \in \{1,\ldots,m\}$, the extrinsic LLR sequence associated with $\mathbf{u}_{t-j,t}$ from the $(t-j)$-th CB $L_{e}^{(t-j)}(\mathbf{u}_{t-j,t})$ becomes the \emph{a priori} LLR sequence $L_a^{(t)}(\mathbf{u}_{t-j,t})$ at the $t$-th CB, the extrinsic LLR sequence associated with $\mathbf{u}_{t,t+j}$ from the $(t+j)$-th CB $L_{e}^{(t+j)}(\mathbf{u}_{t,t+j})$ becomes the \emph{a priori} LLR sequence $L_{a}^{(t)}(\mathbf{u}_{t,t+j})$ at the $t$-th CB as illustrated in Fig. \ref{fig:dec_1}, and the channel LLR $L_{\text{CH}}^{(t-j)}(\mathbf{u}_{t-j,t})$ is from the $(t-j)$-th CB because $\mathbf{u}_{t-j,t}$ is punctured at the $t$-th CB. Due to initialization and termination, $L_{e}^{(t-j)}(\mathbf{u}_{t-j,t})=\infty$ for $t\in \{1,\ldots,m\}, j \in\{ t,\ldots,m\}$ and $L_{e}^{(t+j)}(\mathbf{u}_{t,t+j})=\infty$ for $t\in \{L-m+1,\ldots,L\}, j \in\{ L-t+1,\ldots,m\}$.

Due to the proposed coupling, the extrinsic information from the $(t-j)$-th CB and the extrinsic information from the $(t+j)$-th CB is passed to the $t$-th CB at the same time for $j \in\{1,\ldots,m\}$. In this way, the iterative decoding performance is improved as we will show in the DE analysis in Section \ref{sec:DE}. The decoding terminates when the maximum number of decoding iterations of the FF-FB decoding is reached. Note that for the BEC channel, we can also terminate the decoding when the number of the erased bits (i.e., the number of LLRs being zero) for the current decoding iteration is the same as that for the previous decoding iteration.

The decoding can also be performed by using sliding window decoding. Given a window size $W\geq m+1$, the FF-FB decoding is performed within a window from the $t$-th CB to the $(t+W-1)$-th CB for $t \in \{ 1,\ldots,L-W+1\}$. The update rules for the input LLRs still follow \eqref{eq:LLR1} and \eqref{eq:LLR2}. When the decoder reaches the maximum number of windowed decoding iterations or there are no errors in the codeword, the sliding window decoder outputs the $t$-th CB and then moves to the decoding window from $(t+1)$-th CB to the $(t+W)$-th CB.

\section{Partially Parity-Coupled Turbo Codes}\label{sec:PPC_TC}
In this section, we introduce a new design on partial coupling where a portion of parity digits from a component codeword are coupled to be a part of the information sequence of other component codewords. We refer to this type of codes as partially parity-coupled turbo codes (PPC-TCs). For ease of understanding, we provide an example and an algorithm to show the detailed construction.

\subsection{Encoding of PPC-TCs}
\begin{figure*}[t!]
	\centering
\includegraphics[width=5.0in,clip,keepaspectratio]{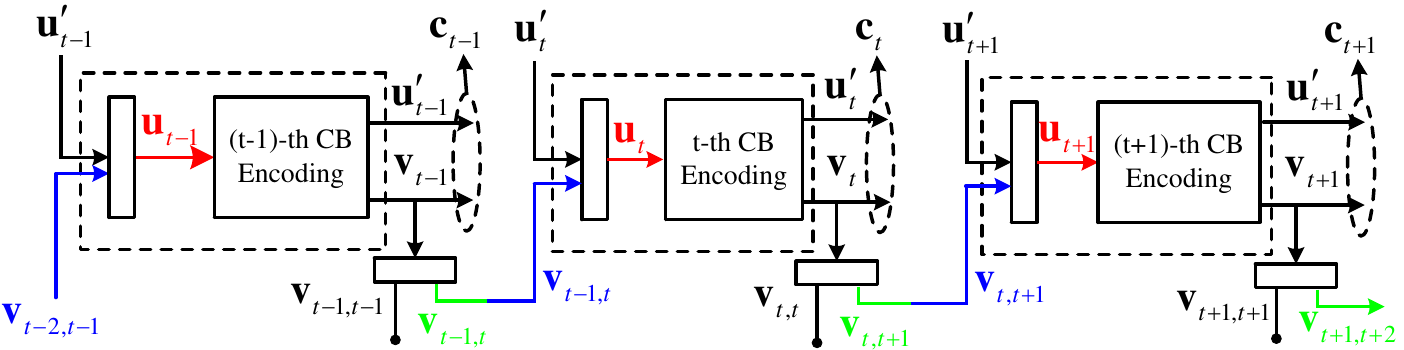}
\caption{Block diagram of the encoder of a PPC-TC with $m=1$.}
\label{fig:ppc1}
\end{figure*}

For ease of explanation, we start with the case of $m=1$. The block diagram of a PPC-TC with $m = 1$ is depicted in Fig. \ref{fig:ppc1}.

Similar to PIC-TCs, we consider that the component code of PPC-TCs is an $(N,K)$ turbo code. First, an information sequence $\mathbf{u}$ is divided into $L$ parts as $\mathbf{u}'_1,\ldots,\mathbf{u}'_L$. Different from PIC-TCs, for the $t$-th CB, $t \in\{1\ldots,L\}$, the message to be encoded is $\mathbf{u}_t = [\mathbf{v}_{t-1,t},\mathbf{u}'_t]$, where $\mathbf{v}_{t-1,t}$ is the coupled parity sequence from the $(t-1)$-th CB and $\mathbf{u}'_t$ is the uncoupled information sequence. Here, the coupled parity sequence is obtained via two steps. As shown in Fig. \ref{fig:ppc1}, the parity sequence of the $(t-1)$-th CB is decomposed into two parts, $\mathbf{v}_{t-1} = [\mathbf{v}_{t-1,t-1}, \mathbf{v}_{t-1,t}]$. Next, $\mathbf{v}_{t-1,t}$ is passed to the $t$-th CB and form the information sequence $\mathbf{u}_t = [\mathbf{v}_{t-1,t},\mathbf{u}'_t]$. After the turbo encoding, the codeword of the $t$-th CB is $\mathbf{c}_t = [\mathbf{u}'_t,\mathbf{v}_{t}]=[\mathbf{u}'_t,\mathbf{v}_{t,t}, \mathbf{v}_{t,t+1}]$. Note that the coupled parity sequence $\mathbf{v}_{t-1,t}$ is punctured in codeword $\mathbf{c}_t$ since it has been included in the codeword for the $(t-1)$-th encoder, i.e., $\mathbf{c}_{t-1} =[\mathbf{u}'_{t-1},\mathbf{v}_{t-1,t-1}, \mathbf{v}_{t-1,t}]$. The coupling length is now $D = n(\mathbf{v}_{t,t+1})$ and the coupling ratio is $\lambda = \frac{D}{K} \in [0,1]$. Then, we have that $n(\mathbf{v}_{t,t+1})=\lambda K$ and $n(\mathbf{v}_{t,t})=N-K-\lambda K$. One can notice that the coupling ratio of PPC-TCs can be naturally extended to any value larger than $\frac{1}{2}$ unlike that of PIC-TCs. For the initialization of  PPC-TCs, zero padding is applied to the first CB such that $\mathbf{v}_{0,1} = \mathbf{0}$. To terminate PPC-TCs, we use an all-zero vector of length $\min\{\lambda K,K-\lambda K\}$ in the $L$-th (last) CB. For $\lambda>\frac{1}{2}$, the number of termination bits is at most $K-\lambda K$. The code rate of a PPC-TC with $m=1$ is
\begin{align}
R_{\text{PPC}} &= \frac{\sum\limits_{t=1}^Ln(\mathbf{u}_t')-\min\{\lambda K,K-\lambda K\}}{\sum\limits_{t=1}^Ln(\mathbf{c}_t)- \min\{\lambda K,K-\lambda K\}} \nonumber \\
&= \frac{L(K-\lambda K)-\min\{\lambda K,K-\lambda K\}}{L(N-\lambda K)- \min\{\lambda K,K-\lambda K\}}.
\end{align}

We use an example to show the construction of PPC-TCs with $m=1$.
\begin{example}
Consider a PPC-TC with $K = 1000$, $L=10$, $\lambda = 1/4$ and $m=1$. Assume that the component code is a rate-$1/3$ turbo code. At time $t=1$, the turbo encoder takes a length $K=1000$ input vector $\mathbf{u}_1 = [\mathbf{v}_{0,1},\mathbf{u}'_1]$, where $\mathbf{v}_{0,1}$ is an all-zero vector of length $\lambda K = 250$ and $\mathbf{u}'_1$ is an information vector of length 750. The component turbo codeword at time $t=1$ is $\mathbf{c}_1 = [\mathbf{u}'_1,\mathbf{v}_{1}]$ of length $N-\lambda K = 2750$, where $\mathbf{v}_{1}$ is the parity sequence of length 2000. Then, at time $t=2$ the parity sequence $\mathbf{v}_{1}$ is decomposed into $\mathbf{v}_{1,1}$ and $\mathbf{v}_{1,2}$, where $\mathbf{v}_{1,2}$ together with the input sequence $\mathbf{u}'_2$ is then encoded by the component turbo encoder. This process continues until $t=10$. The input to the component turbo code becomes $\mathbf{u}_{10} = [\mathbf{v}_{9,10},\mathbf{u}'_{10},\mathbf{0}]$, where $\mathbf{u}'_{10}$ is an information vector of length 500 and $\mathbf{0}$ is a length-250 all-zero vector for termination. The component turbo codeword is $\mathbf{c}_{10}=[\mathbf{u}'_{10},\mathbf{v}_{10}]$ of length 2500. Finally, the codeword of the PPC-TC is $[\mathbf{c}_1,\ldots,\mathbf{c}_{10}] = [\mathbf{u}'_1,\mathbf{v}_{1},\ldots,\mathbf{u}'_{10},\mathbf{v}_{10}]$ of length 27250.
\end{example}

We now consider the case of coupling memory $m \geq 1$. At time $t$, the parity sequence of the $t$-th CB is divided into $m$ parts as $\mathbf{v}_t = [\mathbf{v}_{t,t},\mathbf{v}_{t,t+1},\ldots,\mathbf{v}_{t,t+m}]$, where for $j\in\{1,\ldots,m\},\mathbf{v}_{t,t+j}$ is the coupled parity sequence to be fed to the $(t+j)$-th CB encoder and $n(\mathbf{v}_{t,t+j}) = \frac{\lambda K}{m}$. The input of the $t$-th CB encoder is the length-$K$ vector $\mathbf{u}_t = [\mathbf{v}_{t-m,t},\ldots,\mathbf{v}_{t-1,t},\mathbf{u}'_t]$. After turbo encoding, we obtain the component codeword as $\mathbf{c}_t = [\mathbf{u}'_t,\mathbf{v}_t] = [\mathbf{u}'_t,\mathbf{v}_{t,t},\mathbf{v}_{t,t+1},\ldots,\mathbf{v}_{t,t+m}]$. The detailed encoding procedures are given in Algorithm \ref{alg:en}. The code rate of a PPC-TC with coupling memory $m$ is
\begin{align}
R_{\text{PPC}} &= \frac{\sum\limits_{t=1}^Ln(\mathbf{u}_t')-\sum\limits_{t=L-m+1}^L\min\{(t-L+m)\frac{\lambda K}{m},K-\lambda K\}}{\sum\limits_{t=1}^Ln(\mathbf{c}_t)- \sum\limits_{t=L-m+1}^L\min\{(t-L+m)\frac{\lambda K}{m},K-\lambda K\}} \nonumber \\
&= \frac{L(K-\lambda K)-\sum\limits_{j=1}^m\min\{j\frac{\lambda K}{m},K-\lambda K\}}{L(N-\lambda K)- \sum\limits_{j=1}^m\min\{j\frac{\lambda K}{m},K-\lambda K\}}.
\end{align}

\begin{algorithm}
\DontPrintSemicolon

  \KwInput{Component code $(N,K)$. Coupling ratio $\lambda \in [0,1]$. Coupling memory $m \geq 1$. Coupling length $L \geq 2m$. Information sequences $\mathbf{u}'_1,\ldots,\mathbf{u}'_L$, where $n(\mathbf{u}'_t) = K-\lambda K$ when $t \in \{1,\ldots,L-m\}$ and $n(\mathbf{u}'_t) =\max\{ K(1+(\frac{L-t}{m}-2)\lambda),0\}$ when $t \in \{L-m+1,\ldots,L\}$
}
  \KwOutput{PPC-TC Codeword $[\mathbf{c}_1,\ldots,\mathbf{c}_L]$, where $n(\mathbf{c}_t) =N-\lambda K$ when $t \in \{1,\ldots,L-m\}$ and $n(\mathbf{c}_t) =\max\{ N+(\frac{L-t}{m}-2)\lambda K,N-K\}$ when $t \in \{L-m+1,\ldots,L\}$
  }
 \For{$t=1\to m,j= t \to m$}{
Initialize zero padding $\mathbf{v}_{t-j,t} = \mathbf{0}$, $n(\mathbf{v}_{t-j,t})=\frac{\lambda K}{m}$.
 }
 \For{$t = 1 \to L$}{

\uIf{$t\leq L-m$}{
Construct input vector $\mathbf{u}_t = [\mathbf{v}_{t-m,t},\ldots,\mathbf{v}_{t-1,t},\mathbf{u}'_t]$, $n(\mathbf{u}_t)=K$.
}
\Else{
Initialize termination vector $\mathbf{0}$ of length $\min\{(t-L+m)\frac{\lambda K}{m},K-\lambda K \}$.\\
Construct input vector $\mathbf{u}_t = [\mathbf{v}_{t-m,t},\ldots,\mathbf{v}_{t-1,t},\mathbf{u}'_t, \mathbf{0}]$.\\
}
Encode $\mathbf{u}_t$ to obtain the full codeword $\mathbf{c}'_t =[\mathbf{u}_t,\mathbf{v}_{t}]$, $n(\mathbf{c}'_{t})=N, n(\mathbf{v}_{t})=N-K$.\\
Keep $\mathbf{u}'_t$ and $\mathbf{v}_{t}$ to obtain the final component codeword $\mathbf{c}_t = [\mathbf{u}'_t,\mathbf{v}_{t}]$.\\
Decompose parity sequence $\mathbf{v}_t=[\mathbf{v}_{t,t},\mathbf{v}_{t,t+1},\ldots,\mathbf{v}_{t,t+m}],n(\mathbf{v}_{t,t})=N-K-\lambda K$.
  }
\caption{PPC-TC Encoding}\label{alg:en}
\end{algorithm}

\begin{remark}
PPC-TCs are somewhat similar to Type-I BCCs \cite{8002601}. Indeed, the coupled parity sequence of a component code becomes the information sequence of another component code and the overall code rate after coupling is lower than that of the uncoupled component code. However, contrary to Type-I BCC, for which the coupling is on the convolutional code level and the whole parity sequences are coupled \cite{8002601}, only parts of the parity sequence are coupled in PPC-TCs and the coupling is on the turbo code level. Hence, the encoder and decoder of the component code for PPC-TCs can be kept unchanged. Similar to PIC-TCs, the parity nodes in the code graphs of PPC-TCs have irregular degrees, which will be shown to greatly benefit the decoding threshold.
\end{remark}

\subsection{Decoding of PPC-TCs}
The decoding of PPC-TCs is similar to that of PIC-TCs except for the LLR update rules. Here, we focus on the input LLR updates for the \emph{coupled} parity bits since the input LLR updates for the information bits $\mathbf{u}'_{t}$ and the uncoupled parity bits $\mathbf{v}_{t,t}$ for $t \in \{1,\ldots,L\}$ are the same as that for uncoupled turbo codes. Following the notations and definitions in Section \ref{sec:dec_PIC}, the LLRs of the coupled parity bits input to the $t$-th CB decoder are updated as follows:
\begin{align}
L^{(t)}(\mathbf{v}_{t-j,t}) &= L_{a}^{(t)}(\mathbf{v}_{t-j,t})+ L_{e}^{(t-j)}(\mathbf{v}_{t-j,t})+ L_{\text{CH}}^{(t-j)}(\mathbf{v}_{t-j,t}), \label{eq:LLR1a}\\
L^{(t)}(\mathbf{v}_{t,t+j}) &= L_{e}^{(t)}(\mathbf{v}_{t,t+j})+ L_{e}^{(t+j)}(\mathbf{v}_{t,t+j})+L_{\text{CH}}^{(t)}(\mathbf{v}_{t,t+j}), \label{eq:LLR2a}
\end{align}
Due to the zero padding, $L_{e}^{(t-j)}(\mathbf{v}_{t-j,t})=\infty$ for $t\in \{1,\ldots,m\}, j \in\{ t,\ldots,m\}$. For $t \in \{L-m+1,\ldots,L\}$, the LLRs associated with the termination bits in the $t$-th CB are set to $\infty$.

\section{Graph Model Representation}\label{sec:per_ans}
In this section, we describe the graph model for PIC-TC and PPC-TC ensembles that will be used for deriving the DE equations in Section \ref{sec:DE}. For illustrative purposes, we consider an example with a rate-$1/3$ turbo code as the underlying component code.

\subsection{Graph Model Representation of PIC-TCs}\label{sec:PIC_graph_1}
\begin{figure}[t!]
	\centering
\includegraphics[width=3.42in,clip,keepaspectratio]{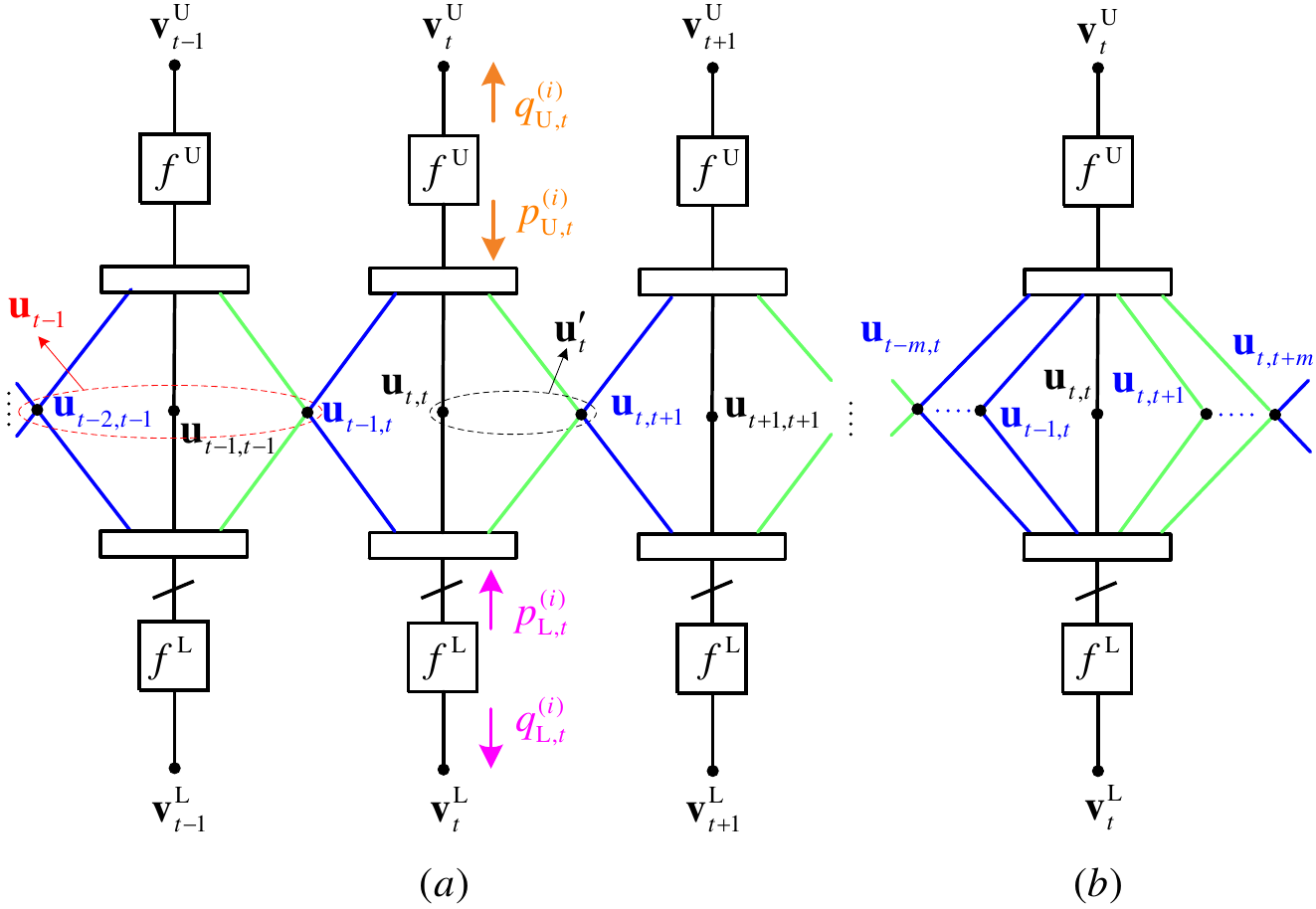}
\caption{Compact graph representation of (a) PIC-TCs with coupling memory $m = 1$ from time instant $t-1$ to $t+1$, and (b) PIC-TCs of coupling memory $m\geq1$ for time instant $t$.}
\label{fig:2}
\end{figure}

We start with the case of PIC-TCs with $m=1$ for ease of explanation. We note that turbo code ensembles can be represented by a compact graph \cite[Sec. III]{8002601}, which simplifies the factor graph representation. The main idea is that each sequences of information bits and parity digits in the factor graph is represented by a single variable node and the trellises are represented by factor nodes. The compact graph representations of PIC-TC ensembles are depicted in Fig. \ref{fig:2}.

Since the component code is a turbo code built from two rate-$\frac{1}{2}$ recursive systematic convolutional codes, the corresponding compact graph of each turbo code has an upper component decoder and a lower component decoder. Two factor nodes $f^{\text{U}}$ and $f^{\text{L}}$ represent the upper and lower convolutional code decoders, respectively, while the decoder takes a length $N$ sequence as its input. The parity sequences associated with the upper and lower decoder at time $t$ are denoted by $\mathbf{v}^{\text{U}}_t$ and $\mathbf{v}^{\text{L}}_t$, respectively. The effect of interleaving on the information sequence of the turbo code is represented by a slash on the edge connecting the information nodes and the lower factor node.

As shown in Fig. \ref{fig:2}(a), for $m=1$ and at time $t$, we use three information nodes to represent $\mathbf{u}_t$\footnote{To ease the notation, we refer to an information node representing a sequence as the sequence itself.} by treating the coupled and uncoupled information sequences separately because these nodes have different degrees. First, the uncoupled information sequence is represented by node $\mathbf{u}_{t,t}$. The coupled information sequence from the decoder at time $t-1$ is represented by node $\mathbf{u}_{t-1,t}$. Here, node $\mathbf{u}_{t-1,t}$ is shared by the compact graphs at time $t-1$ and time $t$, where we recall that the coupled information sequence $\mathbf{u}_{t-1,t}$ is the input of both the $t$-th CB decoder and the $(t-1)$-th CB decoder. It can be seen that the coupled information sequences $\mathbf{u}_{t-1,t}$ is encoded by four convolutional encoders at time $t-1$ and time $t$. Hence, the extrinsic information is passed between \emph{the upper and lower} factor nodes at time $t-1$ and time $t$ \emph{via} node $\mathbf{u}_{t-1,t}$. On the other hand, the uncoupled information node $\mathbf{u}_{t,t}$ only connects the upper and lower factor nodes at time $t$ as these information bits are only encoded by the two constituent convolutional encoders at time $t$. The extrinsic information from $\mathbf{u}_t$ at time $t$ will be propagated to the upper or the lower factor nodes at time instances, $t-1,t$, and $t+1$, simultaneously.

Fig. \ref{fig:2}(b) shows the compact graph representation of the PIC-TC ensemble with coupling memory $m \geq 1$. Due to the partially information coupling, there are $m$ information nodes $\mathbf{u}_{t-m,t},\ldots,\mathbf{u}_{t-1,t}$ connected to $m$ compact graphs from time instances $t-m$ to $t-1$, respectively. Then, there are also $m$ information nodes $\mathbf{u}_{t,t+1},\ldots,\mathbf{u}_{t,t+m}$ which are connected to $m$ compact graphs for time instances $t+1,\ldots,t+m$, respectively. From the $(t+j)$-th decoder perspective, the information node $\mathbf{u}_{t,t+j}$ is shared between the $t$-th compact graph and the $(t+j)$-th compact graph for $j \in \{1,\ldots,m\}$. As a result, the extrinsic information of $\mathbf{u}_{t,t+j}$ is propagated between the upper and lower decoders at time instance $t$ and $t+j$, where the extrinsic information on different edges is \emph{independent} of each other. The uncoupled information node $\mathbf{u}_{t,t}$ only connects the upper and lower factor nodes at time $t$.

\begin{figure}[t!]
	\centering
\includegraphics[width=3.42in,clip,keepaspectratio]{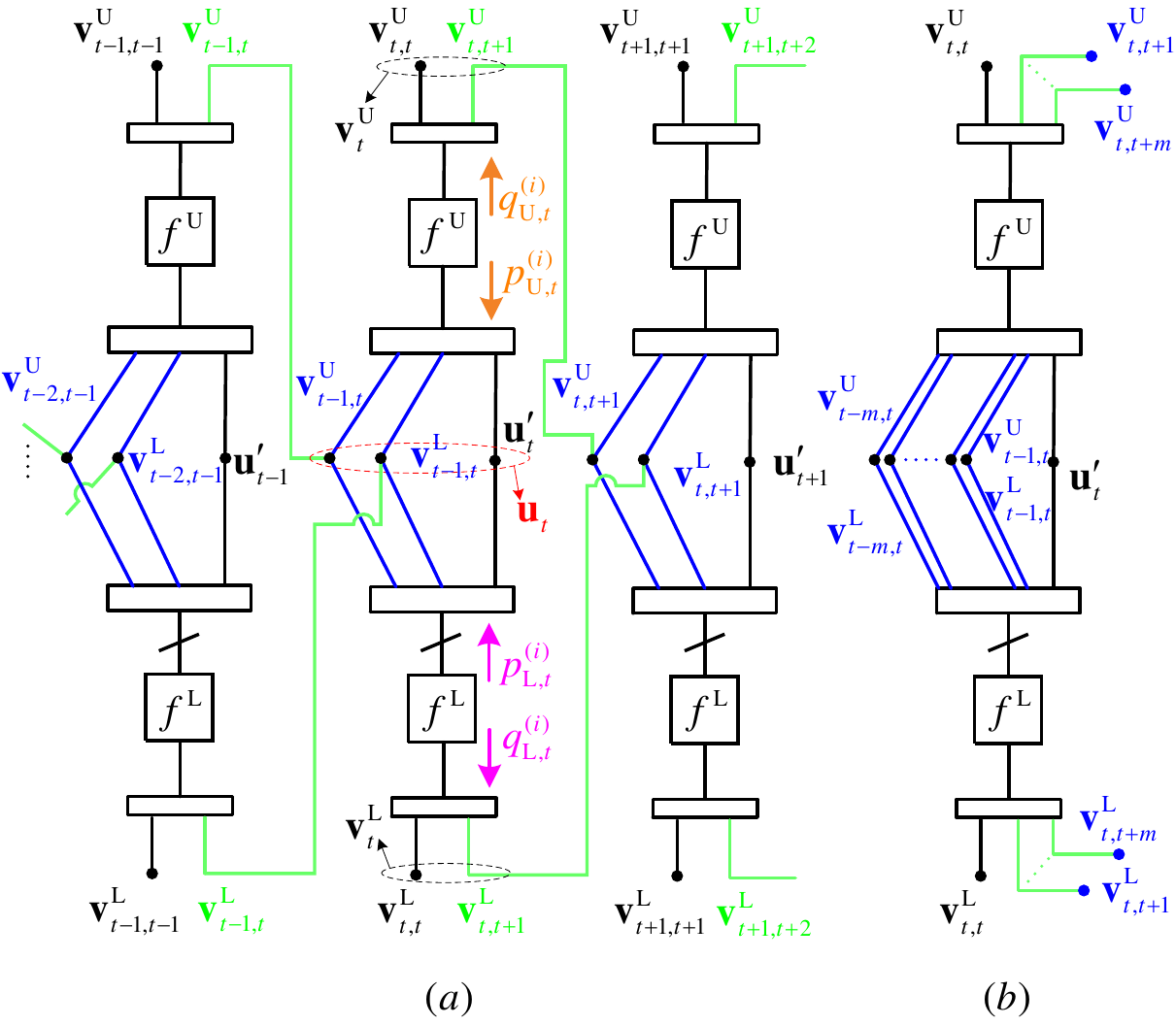}
\caption{Compact graph representation of (a) PPC-TCs with coupling memory $m = 1$ from time instant $t-1$ to $t+1$, and (b) PPC-TCs of coupling memory $m\geq1$ for time instant $t$.}
\label{fig:PC-g}
\end{figure}

\subsection{Graph Model Representation of PPC-TCs}
We start with the case of $m=1$. The compact graph of the PPC-TC for $m=1$ is depicted in Fig. \ref{fig:PC-g}(a).

First, let node $\mathbf{u}'_t$ represent the uncoupled information sequence at time $t$. We also note that the parity sequence consists of the upper and lower parity sequences, i.e., $\mathbf{v}_t = [\mathbf{v}^{\text{U}}_t,\mathbf{v}^{\text{L}}_t]$ for $t \in\{1,\ldots,L\}$. Each parity sequence is further decomposed into $\mathbf{v}^{\text{U}}_t = [\mathbf{v}^{\text{U}}_{t,t},\mathbf{v}^{\text{U}}_{t,t+1}],\mathbf{v}^{\text{L}}_t = [\mathbf{v}^{\text{L}}_{t,t},\mathbf{v}^{\text{L}}_{t,t+1}]$, where $\mathbf{v}^{\text{U}}_{t,t+1}$ and $\mathbf{v}^{\text{L}}_{t,t+1}$ are the coupled parity sequences output from the upper and lower decoders at time $t$, respectively, and $\mathbf{v}^{\text{U}}_{t,t}$ and $\mathbf{v}^{\text{L}}_{t,t}$ are their uncoupled counterparts, respectively. Analogous to PIC-TCs, we refer to the nodes representing the coupled and uncoupled parity sequences as the sequences themselves. The rest of the notations and definitions follow from those of PIC-TCs in Section \ref{sec:PIC_graph_1}.

Different from PIC-TCs, the coupled parity node $\mathbf{v}^{\text{U}}_{t,t+1}$ ($\mathbf{v}^{\text{L}}_{t,t+1}$) at time $t$ is seen as the coupled information node from the ($t+1$)-th decoder perspective. Hence, the extrinsic information of the coupled parity sequences is propagated between the parity output of the upper (lower) decoder at time $t$ and the information output of the upper and lower decoder at time $t+1$ via the shared node $\mathbf{v}^{\text{U}}_{t,t+1}$ ($\mathbf{v}^{\text{L}}_{t,t+1}$). The information node $\mathbf{u}_t$ is a super node consisting of nodes $\mathbf{v}^{\text{U}}_{t-1,t}$, $\mathbf{v}^{\text{L}}_{t-1,t}$, and $\mathbf{u}'_t$. The information flow for the uncoupled parity and information sequences has the same behavior as that for uncoupled turbo codes. For the case of coupling memory $m\geq 1$ shown in Fig. \ref{fig:PC-g}(b), at time $t$ there are $2m$ parity nodes $\mathbf{v}^{\text{L}}_{t,t+j}$ and $\mathbf{v}^{\text{U}}_{t,t+j}$ for $j\in \{1,\ldots,m\}$ that connect the compact graphs at time instances, $t+j$ and $t$.

\section{Density Evolution Analysis for PIC-TCs and PPC-TCs}\label{sec:DE}
In this section, we derive the exact DE equations of PIC-TCs and PPC-TCs based on the graph models introduced in Section \ref{sec:per_ans} for transmissions over the BEC and analyze their decoding thresholds. We denote by $\epsilon$ the channel erasure probability. In what follows, we first show the DE analysis for $m=1$. This simple case will allow us to explain all the important ingredients of our analysis. The derivation for $m\geq1$ for PIC-TCs and PPC-TCs is deferred to Appendix \ref{app:PIC} and Appendix \ref{app:PPC}, respectively.

\subsection{Density Evolution for PIC-TCs}\label{sec:DE_PIC}
For the compact graph at time $t$, we let $p^{(i)}_{\text{U},t}$ and $q^{(i)}_{\text{U},t}$ represent the average extrinsic erasure probability from $f^{\text{U}}$ to $\mathbf{u}_t$ and $\mathbf{v}^{\text{U}}_t$, respectively, after $i$ decoding iterations. Similarly, we let $p^{(i)}_{\text{L},t}$ and $q^{(i)}_{\text{L},t}$ represent the average extrinsic erasure probability from $f^{\text{L}}$ to $\mathbf{u}_t$ and $\mathbf{v}^{\text{L}}_t$, respectively, after $i$ decoding iterations. The transfer function of the upper factor node for information bits and parity bits are denoted by $f^{\text{U}}_p$ and $f^{\text{U}}_q$, respectively. Similarly, the transfer function of the lower factor node for information bits and parity bits are denoted by $f^{\text{L}}_p$ and $f^{\text{L}}_q$, respectively. The transfer function of a specific convolutional code with BCJR decoding \cite{1055186} under the BEC can be explicitly derived by using the methods proposed in \cite{370145,1258535}.

We denote by $\bar{p}^{(i)}_{\text{L},t}$ the average erasure probability of all information nodes, i.e., $\mathbf{u}_t$, input to factor node $f^{\text{U}}$ at time $t$. Based on the graph model in Fig. \ref{fig:2}(a), $\bar{p}^{(i)}_{\text{L},t}$ is the \emph{weighted sum} of the erasure probabilities of information nodes $\mathbf{u}_{t-1,t}$, $\mathbf{u}_{t,t}$, and $\mathbf{u}_{t,t+1}$ to node $f^{\text{U}}$, where the weights are determined by the coupling ratio $\lambda$. Specifically, $\bar{p}^{(i)}_{\text{L},t}$ is
\begin{align}\label{eq:1}
\bar{p}^{(i)}_{\text{L},t}
=& \epsilon\Big(\lambda\cdot p^{(i-1)}_{\text{U},t-1}\cdot p^{(i)}_{\text{L},t-1}\cdot  p^{(i)}_{\text{L},t} +  (1 - 2\lambda) p^{(i)}_{\text{L},t} \nonumber \\
&+  \lambda \cdot  p^{(i)}_{\text{L},t} \cdot p^{(i)}_{\text{L},t+1}\cdot p^{(i-1)}_{\text{U},t+1}\Big),
\end{align}
where $1-2\lambda$ is the ratio of the uncoupled information sequences. Here, $p^{(i-1)}_{\text{U},t-1}\cdot p^{(i)}_{\text{L},t-1}\cdot  p^{(i)}_{\text{L},t}$ is the extrinsic erasure probability on the edge connecting the information node $\mathbf{u}_{t-1,t}$ and factor node $f^{\text{U}}$ at time $t$. It is noteworthy that the multiplication of the erasure probabilities is due to that the decoder can obtain two independent pieces of extrinsic information regarding $\mathbf{u}_{t-1,t}$. This in fact coincides with the addition of LLR sequences for the coupled information sequence as shown in \eqref{eq:LLR1} and \eqref{eq:LLR2} with $m=1$. Likewise, the extrinsic erasure probability on the edge connecting the information node $\mathbf{u}_{t,t+1}$ and $f^{\text{U}}$ is obtained as $p^{(i)}_{\text{L},t} \cdot p^{(i)}_{\text{L},t+1}\cdot p^{(i-1)}_{\text{U},t+1}$ by following \eqref{eq:LLR1} and \eqref{eq:LLR2} with $m=1$.

The DE updates at node $f^{\text{U}}$ at time $t$ are
\begin{align}
p^{(i)}_{\text{U},t} &= f^{\text{U}}_p\left(\bar{p}^{(i)}_{\text{L},t},\bar{q}^{(i)}_{\text{U},t}\right), \label{DE_update_1} \\
q^{(i)}_{\text{U},t} &= f^{\text{U}}_q\left(\bar{p}^{(i)}_{\text{L},t},\bar{q}^{(i)}_{\text{U},t}\right),\label{DE_update_1a}
\end{align}
where $\bar{q}^{(i)}_{\text{U},t} = \epsilon$ is the average erasure probability of $\mathbf{v}^{\text{U}}_t$ to factor node $f^{\text{U}}$ at time $t$.

Likewise, the average erasure probability from $\mathbf{u}_t$ to node $f^{\text{L}}$ at time $t$ is
\begin{align}\label{eq:2}
\bar{p}^{(i)}_{\text{U},t} =& \epsilon\Big(\lambda\cdot p^{(i-1)}_{\text{L},t-1}\cdot p^{(i)}_{\text{U},t-1}\cdot p^{(i)}_{\text{U},t}+  (1 -  2\lambda) p^{(i)}_{\text{U},t}  \nonumber \\
+&  \lambda \cdot p^{(i)}_{\text{U},t}\cdot p^{(i)}_{\text{U},t+1}\cdot p^{(i-1)}_{\text{L},t+1}\Big).
\end{align}
The DE updates at node $f^{\text{L}}$ at time $t$ are
\begin{align}
p^{(i)}_{\text{L},t} = f^{\text{L}}_p\left(\bar{p}^{(i)}_{\text{U},t},\bar{q}^{(i)}_{\text{L},t}\right), \label{DE_update_2} \\
q^{(i)}_{\text{L},t} = f^{\text{L}}_q\left(\bar{p}^{(i)}_{\text{U},t},\bar{q}^{(i)}_{\text{L},t}\right), \label{DE_update_2a}
\end{align}
where $\bar{q}^{(i)}_{\text{L},t} = \epsilon$ is the erasure probability of $\mathbf{v}^{\text{L}}_t$ to factor node $f^{\text{L}}$ at time $t$.

Finally, the \emph{a-posteriori} erasure probability of $\mathbf{u}_t$ at the $i$-th decoding iteration is
\begin{align}
p_{\mathbf{u}_t}^{(i)} = \epsilon \cdot p^{(i)}_{\text{L},t}\cdot p^{(i)}_{\text{U},t}.
\end{align}

\subsection{Density Evolution for PPC-TCs}\label{sec:DE_PPC}
We define the coupling ratio of the upper and lower parity sequences by $\lambda_q^{\text{U}} \triangleq \frac{n(\mathbf{v}^{\text{U}}_{t,t+1})}{K}$ and $\lambda_q^{\text{L}} \triangleq \frac{n(\mathbf{v}^{\text{L}}_{t,t+1})}{K}$, respectively. Then, the overall coupling ration is $\lambda = \lambda_p^{\text{U}}+ \lambda_p^{\text{L}}$.

At time $t$ and iteration $i$, the average extrinsic erasure probability from the super node $\mathbf{u}_t$ to node $f^{\text{U}}$ is
\begin{align}\label{eq:PPC_DE_1}
\bar{p}^{(i)}_{\text{L},t}
 =&\epsilon\Big(\lambda_q^{\text{U}}\cdot q^{(i-1)}_{\text{U},t-1}\cdot  p^{(i)}_{\text{L},t} +  \left(1 - \lambda_q^{\text{U}}-\lambda_q^{\text{L}}\right) p^{(i)}_{\text{L},t} \nonumber \\
 &+ \lambda_q^{\text{L}} \cdot  q^{(i)}_{\text{L},t-1}\cdot  p^{(i)}_{\text{L},t}\Big).
\end{align}
Here, $q^{(i)}_{\text{L},t-1}\cdot p^{(i)}_{\text{L},t}$ is the extrinsic erasure probability on the edge connecting the coupled parity node $\mathbf{v}^{\text{L}}_{t-1,t}$ and the factor node $f^{\text{U}}$ at time $t$. The multiplication is due to the fact that the extrinsic information regarding the parity sequence $\mathbf{v}^{\text{L}}_{t-1,t}$ is obtained independently from node $f_L$ at time $t+1$ and node $\mathbf{u}_t$ at time $t$, which coincides with the LLR updates shown in \eqref{eq:LLR1a} and \eqref{eq:LLR2a}. Likewise, $q^{(i)}_{\text{U},t-1}\cdot  p^{(i)}_{\text{L},t} $ is the extrinsic erasure probability on the edge connecting the coupled parity node $\mathbf{v}^{\text{U}}_{t-1,t}$ and the factor node $f^{\text{U}}$ at time $t$.

Next, the DE updates at node $f^{\text{U}}$ at time $t$ are
\begin{align}
p^{(i)}_{\text{U},t} &= f^{\text{U}}_p\left(\bar{p}^{(i)}_{\text{L},t},\bar{q}^{(i)}_{\text{U},t}\right), \\
q^{(i)}_{\text{U},t} &= f^{\text{U}}_q\left(\bar{p}^{(i)}_{\text{L},t},\bar{q}^{(i)}_{\text{U},t}\right),
\end{align}
where the \emph{weighted sum} of the average erasure probability of $\mathbf{v}^{\text{U}}_{t,t+1}$ to $f^{\text{U}}$ and that of $\mathbf{v}^{\text{U}}_{t,t}$ to $f^{\text{U}}$ is given by
\begin{align}\label{eq:PPC_DE_1a}
\bar{q}^{(i)}_{\text{U},t} = \epsilon\left(\left(1-\lambda_q^{\text{U}}\right)+\lambda_q^{\text{U}} p^{(i)}_{\text{L},t+1}\cdot p^{(i-1)}_{\text{U},t+1}\right).
\end{align}

Similarly, the average erasure probability from the super node $\mathbf{u}_t$ to node $f^{\text{L}}$ at time $t$ is
\begin{align}\label{eq:PPC_DE_2}
\bar{p}^{(i)}_{\text{U},t} =& \epsilon\Big(\lambda_p^{\text{U}}\cdot q^{(i)}_{\text{U},t-1}\cdot p^{(i)}_{\text{U},t} +  \left(1 -  \lambda_p^{\text{U}}-\lambda_p^{\text{L}}\right) p^{(i)}_{\text{U},t} \nonumber \\
&+  \lambda_p^{\text{L}} \cdot p^{(i)}_{\text{U},t}\cdot q^{(i)}_{\text{L},t-1}\Big),
\end{align}
and DE updates at node $f^{\text{L}}$ at time $t$ are
\begin{align}
p^{(i)}_{\text{L},t} = f^{\text{L}}_p\left(\bar{p}^{(i)}_{\text{U},t},\bar{q}^{(i)}_{\text{L},t}\right), \\
q^{(i)}_{\text{L},t} = f^{\text{L}}_q\left(\bar{p}^{(i)}_{\text{U},t},\bar{q}^{(i)}_{\text{L},t}\right),
\end{align}
where the \emph{weighted sum} of the average erasure probability of $\mathbf{v}^{\text{L}}_{t,t+1}$ to $f^{\text{L}}$ and that of $\mathbf{v}^{\text{L}}_{t,t}$ to $f^{\text{L}}$ is given by
\begin{align}\label{eq:PPC_DE_2a}
\bar{q}^{(i)}_{\text{L},t} = \epsilon((1-\lambda_p^{\text{L}})+\lambda_p^{\text{L}} p^{(i)}_{\text{U},t}\cdot p^{(i-1)}_{\text{L},t+1}).
\end{align}

Finally, the \emph{a-posteriori} erasure probability of $\mathbf{u}_t$ at the $i$-th decoding iteration is
\begin{align}
p_{\mathbf{u}_t}^{(i)} = \epsilon \cdot p^{(i)}_{\text{L},t}\cdot p^{(i)}_{\text{U},t}.
\end{align}

\begin{table}[t!]
  \centering
 \caption{Thresholds For PIC-TCs and PPC-TCs}\label{table_PIC_m}
\begin{tabular}{c c c c c c c}
\hline
  Ensemble &Rate  & $\lambda$   & $\epsilon^{(m=1)}_{\text{BP}}$  & $\epsilon^{(m=15)}_{\text{BP}}$  & $\delta_{\epsilon}^{(m=1)}$ & $\delta_{\epsilon}^{(m=15)}$ \\  \hline
 PIC-TC & 0.3191 & 1/16  & 0.6628  & 0.6633  & 0.0181& 0.0176 \\
 PPC-TC & 0.3191 & 1/16   & 0.6622 & 0.6628 &0.0187 & 0.0181\\
 \hline
  PIC-TC &  0.3043 & 1/8  & 0.6818 & 0.6828  & 0.0139& 0.0127\\
  PPC-TC &  0.3043 & 1/8  & 0.6803 &0.6818 & 0.0154& 0.0139\\
  \hline
  PIC-TC & 0.3000 & 1/7   &  0.6871  & 0.6883  & 0.0129& 0.0017 \\
  PPC-TC & 0.3000& 1/7   & 0.6854 & 0.6872  & 0.0146 & 0.0028\\
  \hline
 PIC-TC &   0.2727 & 1/4   & 0.7182 & 0.7206  & 0.0091& 0.0067\\
 PPC-TC &  0.2727 & 1/4   & 0.7156 &  0.7190 &0.0117  & 0.0070\\
  \hline
  PIC-TC &  0.2500 & 1/3   & 0.7424 & 0.7457  & 0.0076 & 0.0043\\
  PPC-TC &  0.2500 & 1/3   & 0.7394 & 0.7441  & 0.0106& 0.0047\\
 \hline
  PIC-TC &  0.2381 & 3/8   & 0.7546  & 0.7585  & 0.0073& 0.0034 \\
   PPC-TC & 0.2381 & 3/8   & 0.7516 &  0.7570 & 0.0103& 0.0039\\
  \hline
  PIC-TC &  0.2000 & 1/2   & 0.7926 & 0.7982  & 0.0074& 0.0018\\
   PPC-TC &  0.2000 & 1/2   & 0.7899 & 0.7970 &0.0101  &0.0030 \\
  \hline
\end{tabular}
\end{table}

\subsection{Density Evolution Results}
In this section, we evaluate the iterative decoding thresholds of PIC-TC and PPC-TC ensembles over the BEC by using the derived DE equations. We consider that the underlying component code is a rate-$1/3$ turbo code with identical 4-state rate-$1/2$ component encoders with generator polynomial $(1,\frac{5}{7})$ in octal notation. As a result, the transfer functions for the upper and lower decoders are the same, i.e., $f^{\text{U}}_p = f^{\text{L}}_p$ and $f^{\text{U}}_q = f^{\text{L}}_q$. For a given coupling memory $m$, the iterative decoding thresholds (denoted by $\epsilon^{(m)}_{\text{BP}}$) and the gap to the BEC capacity (denoted by $\delta_{\epsilon}^{(m)}$) for PIC-TCs and PPC-TCs with various values of $\lambda$ are shown in Table \ref{table_PIC_m}. For the PPC-TCs, we choose $\lambda^{\text{U}}_p = \lambda^{\text{L}}_p$ as we observed that this choice yields the best threshold.

\begin{figure}[t!]
	\centering
\includegraphics[width=3.42in,clip,keepaspectratio]{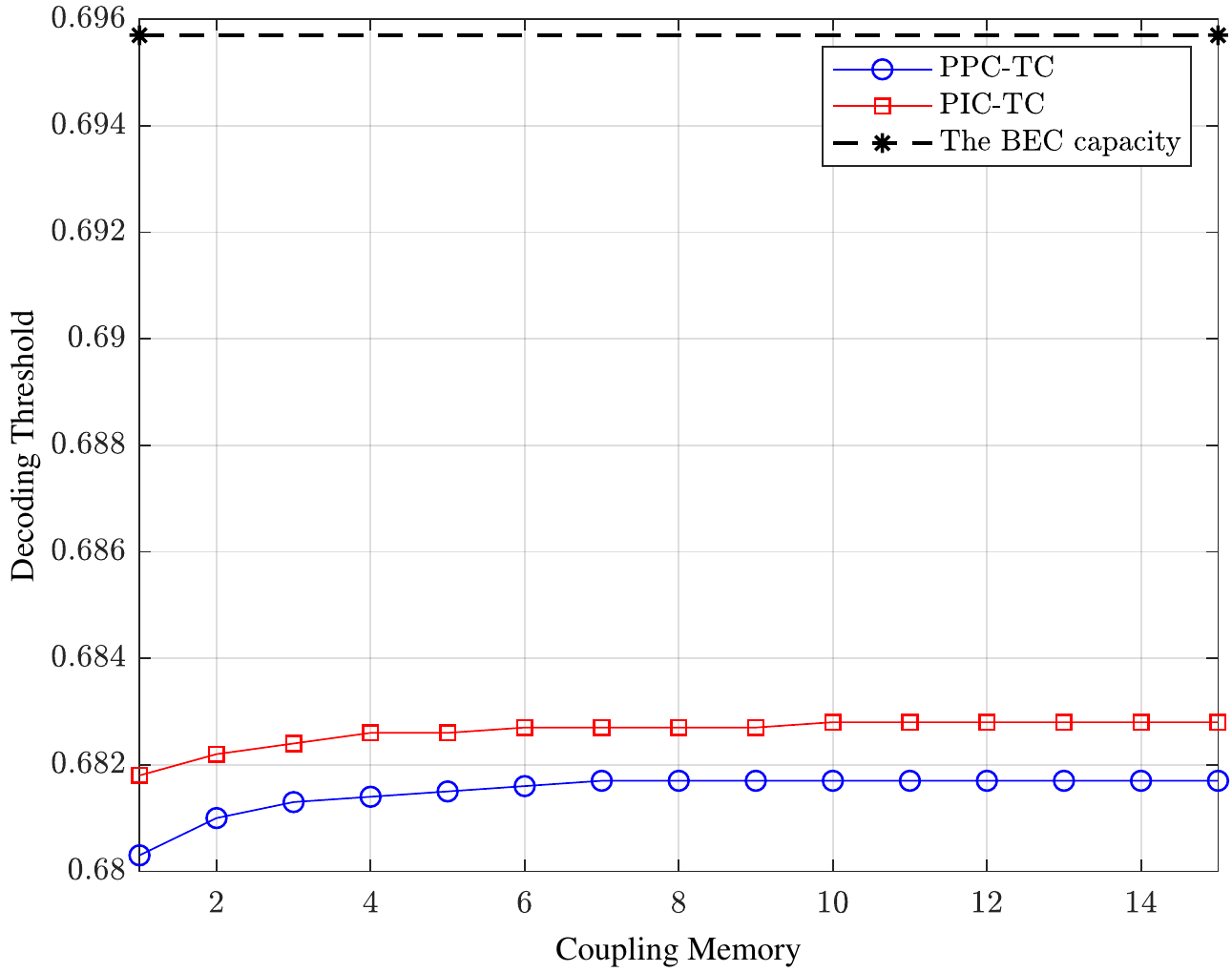}
\caption{Thresholds versus the coupling memory with $\lambda = 1/8$.}
\label{fig:cpm0p3034}
\end{figure}
\begin{figure}[t!]
	\centering
\includegraphics[width=3.42in,clip,keepaspectratio]{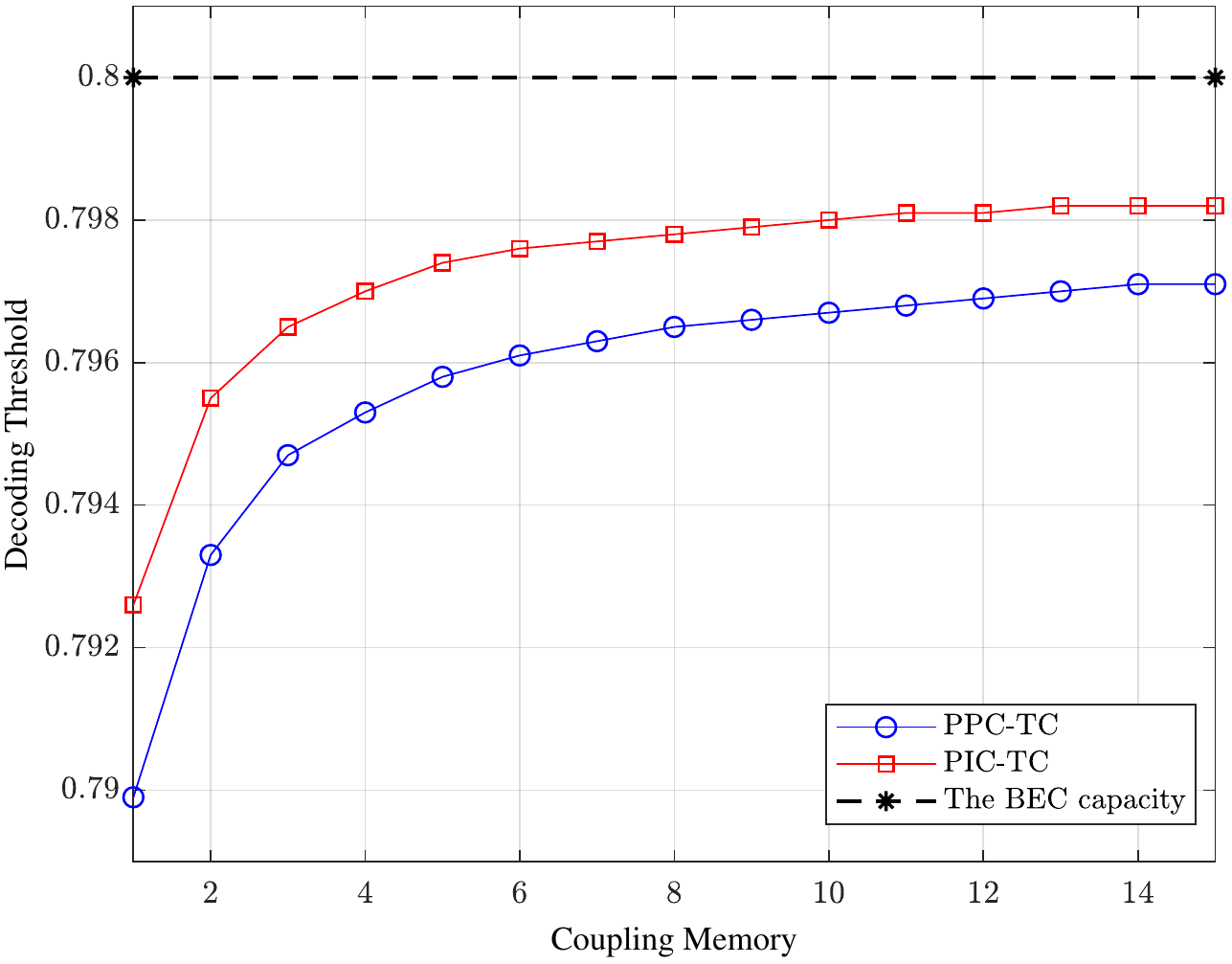}
\caption{Thresholds versus the coupling memory with $\lambda= 1/2$.}
\label{fig:fig:cpm0p2}
\end{figure}

From Table \ref{table_PIC_m}, it can be seen that the DE thresholds of both PIC-TCs and PPC-TCs are close to the BEC capacity and PIC-TCs perform better than PPC-TCs for the same coupling ratio and coupling memory. Further, the gap between the threshold and the BEC capacity decreases with increasing $\lambda$ and the thresholds improve with increasing coupling memory. It is interesting to note that the threshold improvements with $m$ are larger for larger coupling ratios. A possible reason could be that since the codes with larger coupling ratios are stronger than those with smaller coupling ratios, the stronger codes benefit significantly from increasing the coupling memory. This phenomena was also observed in \cite{8002601}, where SC-SCCs have much larger decoding threshold improvements over SC-PCCs with increasing $m$ because the MAP threshold of uncoupled SCCs is much better than that of uncoupled PCCs.

To see the details on how the thresholds behave when we increase the coupling memory, we plot the decoding thresholds of PIC-TCs and PPC-TCs versus the coupling memory in Fig. \ref{fig:cpm0p3034} and
Fig. \ref{fig:fig:cpm0p2} for $\lambda = 1/8$ and $1/2$, respectively.

From Figs. \ref{fig:cpm0p3034}-\ref{fig:fig:cpm0p2}, we can see that the decoding thresholds of PIC-TC ensembles perform closer to the BEC capacity than PPC-TC ensembles for the same coupling memory and coupling ratio. It is also observed that the improvement of the decoding threshold is fast when the coupling memory $m$ is increased from 1 to 5 and this improvement becomes slow after $m>5$. Although it is unclear what are decoding thresholds of PIC-TC and PPC-TC as $m \rightarrow \infty$, the thresholds of our codes are still increasing when $m>15$ as we will see in Section \ref{sec:RC_PIC_1}.

\subsection{Error Performance Simulation and Comparison}
We present the numerical results on the error performance for PIC-TCs and PPC-TCs with $m=1$ under the FF-FB decoding over the BEC. Unless specified otherwise, we use random interleavers and randomly choose the positions of coupled bits (random for each transmission).

\begin{figure}[t!]
	\centering
\includegraphics[width=3.42in,clip,keepaspectratio]{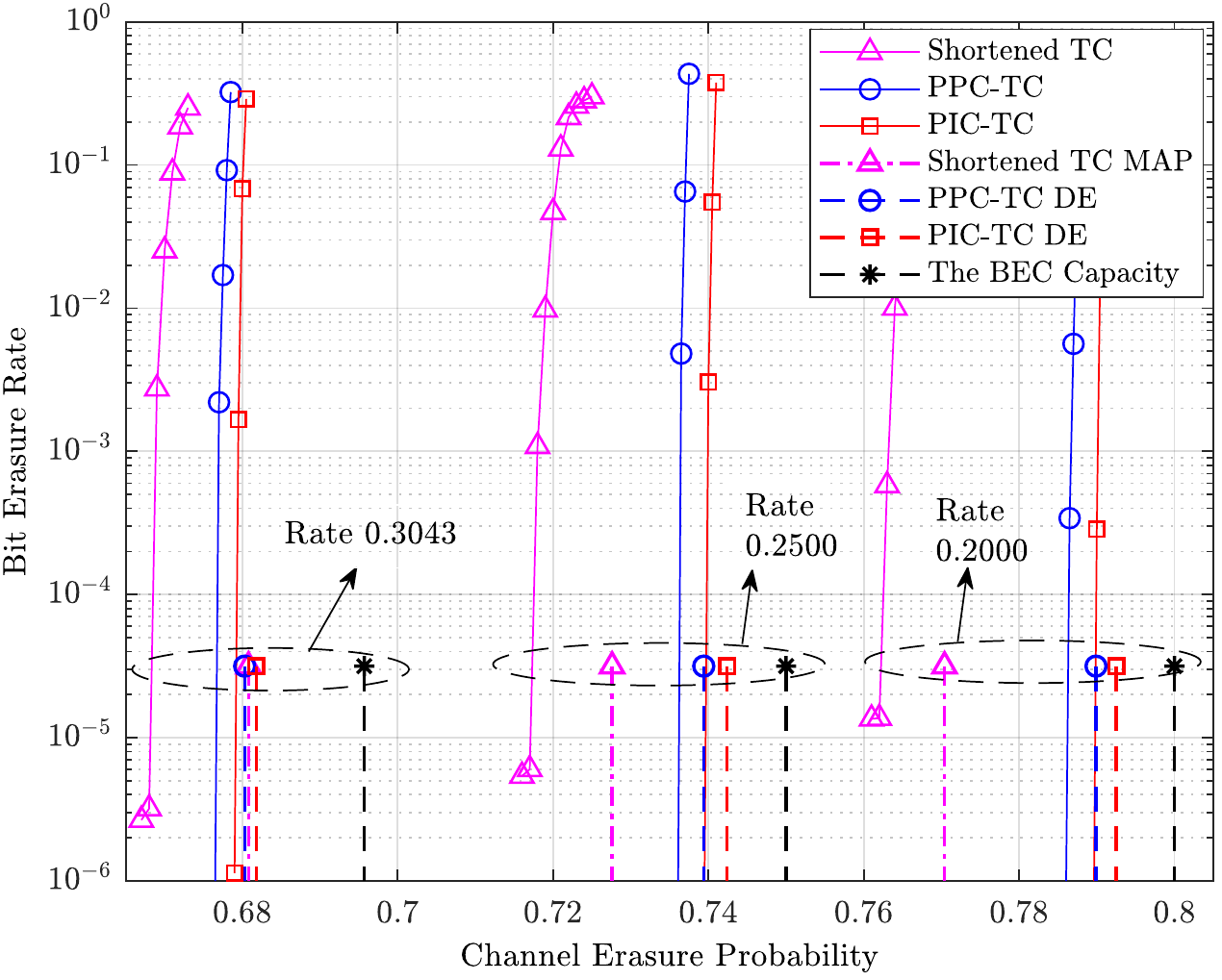}
\caption{Bit erasure rate of several PIC-TCs, PPC-TCs and shortened TCs with different code rates.}
\label{fig:PIC_PPC_BER}
\end{figure}

Recall that the component turbo code of PIC-TCs and PPC-TCs is a rate-$1/3$ turbo code. To see the erasure rate performance for large block length, we set the information length of the component code $K = 100000$. In addition, we set $L = 100$ to minimize the code rate loss due to zero padding. We consider $\lambda = 1/8, 1/3$, and $1/2$, corresponding to the code rates of 0.3043, 0.2500, and 0.2000, respectively. The performance is measured in terms of bit erasure rate (BER) versus the erasure probability of the BEC and is shown in Fig. \ref{fig:PIC_PPC_BER}. Moreover, the corresponding DE thresholds are included in the figure.

To put the results of the proposed scheme in context, we also include the error performance and the MAP decoding thresholds of shortened turbo codes in Fig. \ref{fig:PIC_PPC_BER}. The MAP thresholds $\epsilon_{\text{MAP}}$ are obtained from the following equation \cite{1523540}
\begin{align}
R = \int_{\epsilon_{\text{MAP}}}^1 p(\epsilon)+(1-R)q(\epsilon) d \epsilon,
\end{align}
where $R$ is the target code rate, $p(\epsilon)$ and $q(\epsilon)$ are the average extrinsic erasure probability for information bits and parity bits, respectively. The values of the MAP thresholds for shortened turbo codes with rates 0.3043, 0.2500 and 0.2000 are obtained as 0.6808 0.7276 and 0.7704, respectively. The simulated shortened turbo codes assume random shortening patterns and with information length 100000. Note that the MAP thresholds of the shortened turbo codes are in fact the iterative decoding threshold of the shortened SC-PCCs \cite{8368318} as it can be easily shown that threshold saturation occurs with shortening.

It can be observed from Fig. \ref{fig:PIC_PPC_BER} that the actual decoding performance of all PIC-TCs and PPC-TCs is in agreement with the corresponding DE thresholds. Moreover, both PIC-TCs and PPC-TCs significantly outperform the shortened turbo codes for all the code rates we have simulated. It is also worth noting that PIC-TCs and PPC-TCs perform significantly better than the shortened SC-PCCs when the coupling ratio is large. This implies that by simply increasing the coupling ratio to lower the code rate and without explicitly designing the code structures, PIC-TCs and PPC-TCs can provide superior performance over shortened turbo codes and shortened SC-PCCs while and approach the BEC capacity.

\section{Rate-Compatible Partially Coupled Turbo Codes}\label{sec:RC_PIC}
In this section, we consider rate-compatible PIC-TCs and PPC-TCs and analyze their decoding thresholds. To obtain higher code rates, we consider applying random puncturing \cite{7353121} on parity bits. To obtain lower code rates, we can further increase the coupling ratio such that $\lambda >1/2$ for PPC-TCs. We also find the best coupling ratio and puncturing ratio pair in order to obtain the optimal decoding threshold for a given code rate and coupling memory.

\subsection{High Rate Performance Under Random Puncturing}\label{sec:RC_PIC_1}
We let $\rho\in [0,1]$ represent the fraction of surviving bits after puncturing. For such a randomly punctured code sequence transmitted over a BEC with erasure probability $\epsilon$, the erasure probability for the code sequence is $\epsilon_{\rho} = 1-(1-\epsilon)\rho$ \cite{8368318}.

For the punctured PIC-TCs, the code rate in \eqref{eq:PICrate3} becomes
\begin{align}\label{eq:rate_punctuured_L}
R_{\text{PIC}} =\frac{1-\lambda}{(\frac{1}{R_0}-1)\rho+1-\lambda }.
\end{align}
Hence, for a given target code rate $R_{\text{PIC}}$ and coupling ratio $\lambda$, we can determine $\rho$ from \eqref{eq:rate_punctuured_L}. The DE equations for the punctured PIC-TCs are obtained by substituting $\epsilon_{\rho} \rightarrow \bar{q}^{(i)}_{\text{U},t}$ in \eqref{DE_update_1} and \eqref{DE_update_1a} and substituting $\epsilon_{\rho} \rightarrow \bar{q}^{(i)}_{\text{L},t}$ in \eqref{DE_update_2} and \eqref{DE_update_2a}.

For PPC-TCs, we have the same code rate as that of PIC-TCs in \eqref{eq:rate_punctuured_L} when puncturing the parity bits. The DE equations for the punctured PPC-TCs requires to modify \eqref{eq:PPC_DE_m_1} and \eqref{eq:PPC_DE_m_2} to
\begin{align}
\bar{p}^{(i)}_{\text{L},t} =&   p^{(i)}_{\text{L},t}\Bigg(\epsilon_{\rho}\sum\limits_{j=1}^m \left(\frac{\lambda_q^{\text{U}}}{m} \cdot q^{(i-1)}_{\text{U},t-j}  + \frac{\lambda_q^{\text{L}}}{m} \cdot q^{(i)}_{\text{L},t-j}\right) \nonumber \\
&+ \epsilon(1 - \lambda_q^{\text{U}}-\lambda_q^{\text{L}}) \Bigg),
\end{align}
and
\begin{align}
\bar{p}^{(i)}_{\text{U},t} =& p^{(i)}_{\text{U},t}\Bigg(\epsilon_{\rho}\sum\limits_{j=1}^m \left(\frac{\lambda_p^{\text{U}}}{m}\cdot q^{(i)}_{\text{U},t-1} +\frac{\lambda_p^{\text{L}}}{m}\cdot q^{(i-1)}_{\text{L},t-1}\right) \nonumber \\
&+  \epsilon (1 -  \lambda_p^{\text{U}}-\lambda_p^{\text{L}} )\Bigg),
\end{align}
respectively, and substitute $\epsilon_{\rho} \rightarrow \epsilon$ in \eqref{eq:PPC_DE_m_1a} and \eqref{eq:PPC_DE_m_2a}.

For both PIC-TCs and PPC-TCs, from \eqref{eq:rate_punctuured_L} it results that for given a code rate $R$, each choice of $\lambda$ corresponds to a unique $\rho$,
\begin{align}
\rho =\frac{\frac{1}{R}-1}{\frac{1}{R_0}-1}(1-\lambda).
\end{align}
For a target code rate and coupling memory, we jointly optimize the pair $(\lambda,\rho)$ such that the iterative decoding thresholds are maximized. The optimized pair $(\lambda,\rho)$ and the corresponding thresholds for PIC-TCs and PPC-TCs for $m = 1,15,50, 200,1000$ are reported in Table \ref{RC_compare_2} and Table \ref{RC_compare_1}, respectively. Note that for some cases, the optimal pair $(\lambda,\rho)$ is given as two sets of values. This is because these choices yield almost the same decoding threshold with only a very minor difference, e.g., $\pm 0.00003$. Note that when $m=1000$, we choose the largest value of $\lambda$ from the case of $m=200$ for simplicity. For comparison purposes, we also report the decoding thresholds for SC-PCCs, SC-SCCs, SC-BCCs (Type-II) \cite[Tabel II]{8368318}.

\begin{table*}[t!]
  \centering
 \caption{Optimal $(\lambda,\rho)$ for Rate Compatible PIC-TCs and PPC-TCs}\label{RC_compare_2}
\begin{tabular}{c c c c c c c}
\hline
Ensemble & Rate     & $m=1$ & $m=15$ &$m=50$ &$m=200$  \\
 \hline
PIC-TC & 9/10	&		  	   (0.5,0.0278) &  (0.5,0.0278) & (0.5,0.0278) & (0.5,0.0278) \\
PPC-TC &9/10	&  ([0.19,0.2],[0.045,0.0444]) &  ([0.43,0.47],[0.0322,0.0289]) & ([0.5,0.64],[0.0316,0.0296])  & ([0.53,0.83],[0.0261,0.0094])  \\
  \hline
 PIC-TC& 4/5	&			(0.5,0.0625) &  (0.5,0.0625)&  (0.5,0.0625) &  (0.5,0.0625)  \\
PPC-TC &4/5	& ([0.25,0.28],[0.0938,0.09])  & ([0.51,0.58],[0.0613,0.0525])  & ([0.59,0.73],[0.0513,0.0338])  & ([0.63,0.88],[0.0463,0.015])  \\
  \hline
 PIC-TC &  3/4		      &(0.5,0.0833) &  (0.5,0.0833) & (0.5,0.0833)  & (0.5,0.0833)  \\
 PPC-TC &3/4	& ([0.28,0.30],([0.12,0.1167])  &  ([0.54,0.61],[0.0767,0.065]) & ([0.61,0.76],[0.065,0.04]) & ([0.69,0.88],[0.0517,0.02]) \\
   \hline
PIC-TC  &  2/3	  &(0.5,0.1250) & (0.5,0.1250)&    (0.5,0.1250) &    (0.5,0.1250)   \\
PPC-TC &2/3	&  ([0.30,0.31],[0.175,0.1725]) &  ([0.56,0.67],[0.11,0.0825]) & ([0.65,0.8],[0.0875,0.05]) & ([0.75,0.9],[0.0625,0.025])  \\
  \hline
 PIC-TC& 1/2	&	    ([0.44,0.48],[0.28,0.26]) &  (0.5,0.25)&  (0.5,0.25)  &  (0.5,0.25)  \\
 PPC-TC &1/2	&  ([0.32,0.33],[0.34,0.335]) &  ([0.64,0.72],[0.18,0.14]) &  ([0.76,0.83],[0.12,0.085]) &  ([0.8,0.93],[0.1,0.035]) \\
   \hline
PIC-TC &   1/3	      &([0.37,0.42],[0.63,0.58]) &(0.5,0.5)& (0.5,0.5)  & (0.5,0.5)  \\
PPC-TC &1/3	& ([0.32,0.35],[0.68,0.65])  & ([0.69,0.78],[0.31,0.22])  & ([0.78,0.88],[0.22,0.12]) & ([0.85,0.95],[0.15,0.05])  \\
  \hline
\end{tabular}
\end{table*}

\begin{table*}[t!]
  \centering
 \caption{thresholds for PIC-TCs, PPC-TCs, SC-PCCs, SC-SCCs and SC-BCCs}\label{RC_compare_1}
\begin{tabular}{c c c c c c c}
\hline
Ensemble & Rate     & $\epsilon^{(m=1)}_{\text{BP}}$ & $\epsilon^{(m=15)}_{\text{BP}}$ &$\epsilon^{(m=50)}_{\text{BP}}$ & $\epsilon^{(m=200)}_{\text{BP}}$ &$\epsilon^{(m=1000)}_{\text{BP}}$  \\
 \hline
PIC-TC & 9/10	&	0.0863 &  0.0881 & 0.0881 & 0.0882 & 0.0882 \\
PPC-TC &9/10	&  0.0931 &  0.0990 & 0.0996 & 0.0998 & 0.0999  \\
SC-PCC &9/10	&  0.0582 &  0.0582 & 0.0582 & 0.0582 & 0.0582  \\
SC-SCC &9/10	&  0.0624 &  0.0996 & 0.0996 & 0.0996 & 0.0996  \\
SC-BCC &9/10	&  0.0954 &  0.0990 & 0.0990 & 0.0990 & 0.0990  \\
  \hline
 PIC-TC& 4/5	&	0.1811 &  0.1846&  0.1847 &  0.1848 & 0.1848  \\
PPC-TC &4/5	& 0.1896  & 0.1985  & 0.1994 & 0.1997 & 0.1999  \\
SC-PCC &4/5	& 0.1391 &  0.1391 & 0.1391  & 0.1391 & 0.1391  \\
SC-SCC &4/5	&  0.1644 & 0.1990 & 0.1990  & 0.1990 & 0.1990 \\
SC-BCC &4/5	& 0.1986 &  0.1999 & 0.1999  & 0.1999 & 0.1999 \\
  \hline
 PIC-TC &  3/4 &0.2307 & 0.2349 & 0.2351 & 0.2351 &  0.2351  \\
 PPC-TC &3/4	& 0.2385  & 0.2483 & 0.2493  &  0.2497&  0.2498 \\
 SC-PCC &3/4& 0.1876 &  0.1876 & 0.1876  & 0.1876 & 0.1876 \\
SC-SCC &3/4	&  0.2155 &  0.2486 & 0.2486  & 0.2486 & 0.2486 \\
SC-BCC &3/4	&  0.2481 &  0.2491 & 0.2491  & 0.2491 & 0.2491 \\
   \hline
PIC-TC  &  2/3	  &0.3151 & 0.3204&    0.3206  &    0.3207 &  0.3207  \\
PPC-TC &2/3	&  0.3206 &  0.3313 & 0.3325  &  0.3330 &  0.3332\\
SC-PCC &2/3	&  0.2772 &  0.2772 & 0.2772  & 0.2772  & 0.2772\\
SC-SCC &2/3	&  0.3303&  0.3316 & 0.3316   & 0.3316  & 0.3316 \\
SC-BCC &2/3	&  0.3323 &  0.3331 & 0.3331  & 0.3331 & 0.3331 \\
  \hline
 PIC-TC& 1/2	&	    0.4865 &  0.4930&  0.4933 &  0.4934 & 0.4934  \\
 PPC-TC &1/2	&  0.4865 &  0.4977 &  0.4991 &  0.4996 &  0.4998\\
 SC-PCC &1/2	&  0.4689&  0.4689 & 0.4689  & 0.4689 & 0.4689 \\
SC-SCC &1/2	&  0.4708 &  0.4981 & 0.4981  & 0.4981 & 0.4981 \\
SC-BCC &1/2	&  0.4988 &  0.4993 & 0.4993  & 0.4993 & 0.4993 \\
   \hline
PIC-TC &   1/3	      &0.6576 &0.6635& 0.6639 & 0.6640 &  0.6640  \\
PPC-TC &1/3	& 0.6545  & 0.6645  & 0.6658  &  0.6664 &  0.6665 \\
SC-PCC &1/3	& 0.6553 &  0.6553 & 0.6553 & 0.6553 & 0.6553 \\
SC-SCC &1/3	&  0.6437 &  0.6654 & 0.6654  & 0.6654 & 0.6654 \\
SC-BCC &1/3	&  0.6651&  0.6653 & 0.6653  & 0.6653 & 0.6653 \\
  \hline
\end{tabular}
\end{table*}

From Table \ref{RC_compare_1}, it can be seen that the decoding thresholds of PIC-TCs and PPC-TCs are better than those of SC-PCCs \cite[Tabel II]{8368318} for the same rates. Satisfactory decoding performance can be attained for both codes with coupling memory $m=1$ and the decoding thresholds of both codes improve with increasing coupling memory. Moreover, PPC-TCs outperform PIC-TCs with random puncturing on the parity bits. The reason is that the coupled bit nodes with puncturing can still provide reliable information due to their larger degrees compared to the uncoupled bit nodes. Furthermore, PPC-TCs perform better than SC-SCCs \cite{8368318} and have comparable performance to SC-BCCs \cite{8368318} when the coupling memory is large. For example, the PPC-TC for rate-$4/5$ has a decoding threshold of 0.1999 while the decoding thresholds for SC-SCCs and SC-BCCs are 0.1990 and 0.1999, respectively, for $m=1000$. Our results strongly suggest that PPC-TCs may approach the BEC capacity asymptotically as the coupling memory tends to infinity. It is also interesting to note in Table \ref{RC_compare_2} that for PIC-TCs the coupling ratio needs to be very large in order to obtain more coupling gain when the code rate is higher or the coupling memory is larger while for PPC-TCs the value of the optimal coupling ratio decreases with the target code rate increased. We also observe that the optimal pair $(\lambda,\rho)$ for PPC-TCs can be chosen from a range of values because these choices lead to the same decoding threshold to 4 decimal place. Therefore, the choice of $(\lambda,\rho)$ for PPC-TCs is even more flexible than that for PIC-TCs. Our results show that by carefully choosing $(\lambda,\rho)$, PIC-TCs and PPC-TCs can operate very close to the BEC capacity for a wide range of code rate without changing the component codes.

\begin{figure}[t!]
	\centering
\includegraphics[width=3.42in,clip,keepaspectratio]{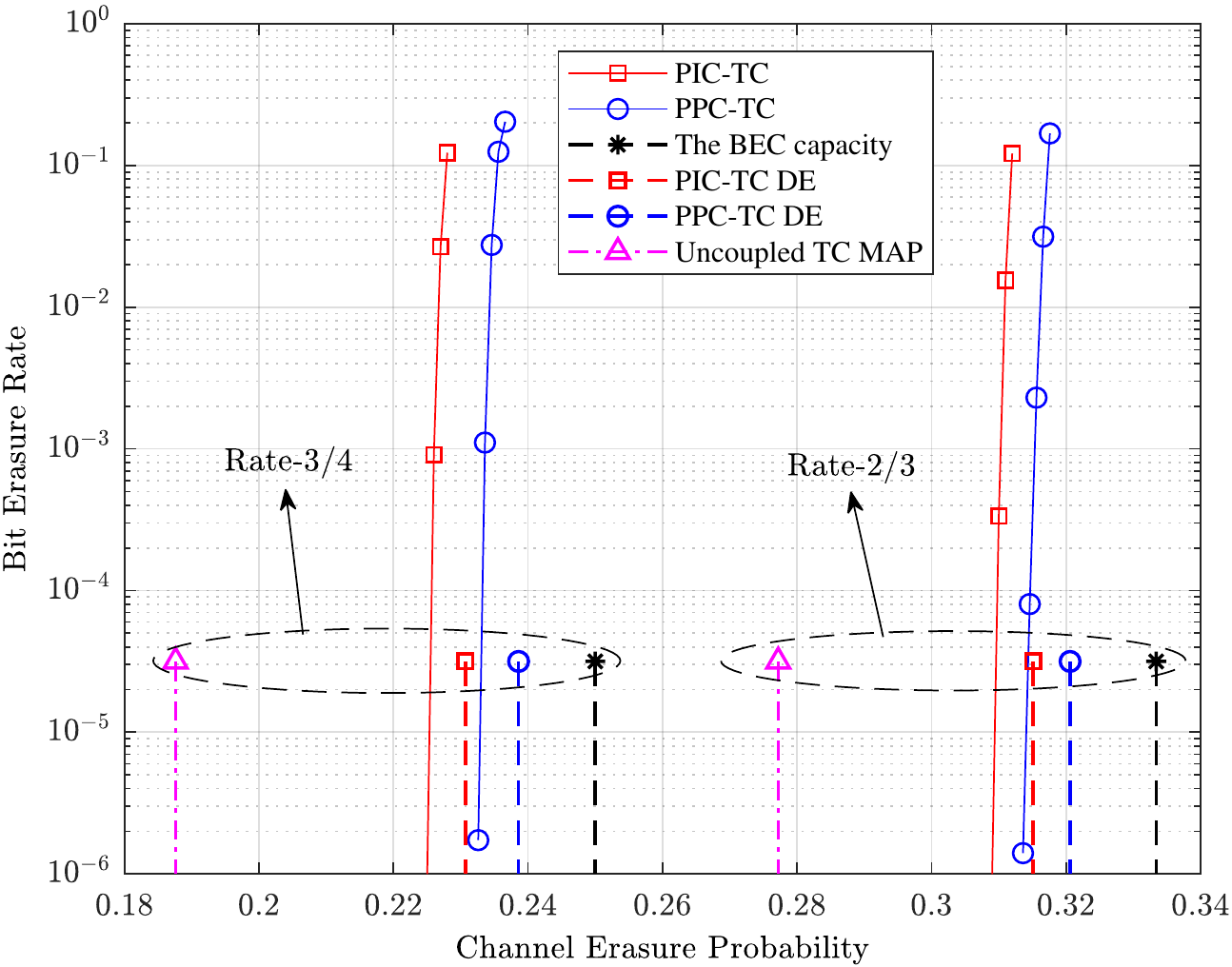}
\caption{Bit erasure rate of PIC-TCs, PPC-TCs with random parity puncturing.}
\label{fig:PIC_punc_rand}
\end{figure}
\begin{figure}[t!]
	\centering
\includegraphics[width=3.42in,clip,keepaspectratio]{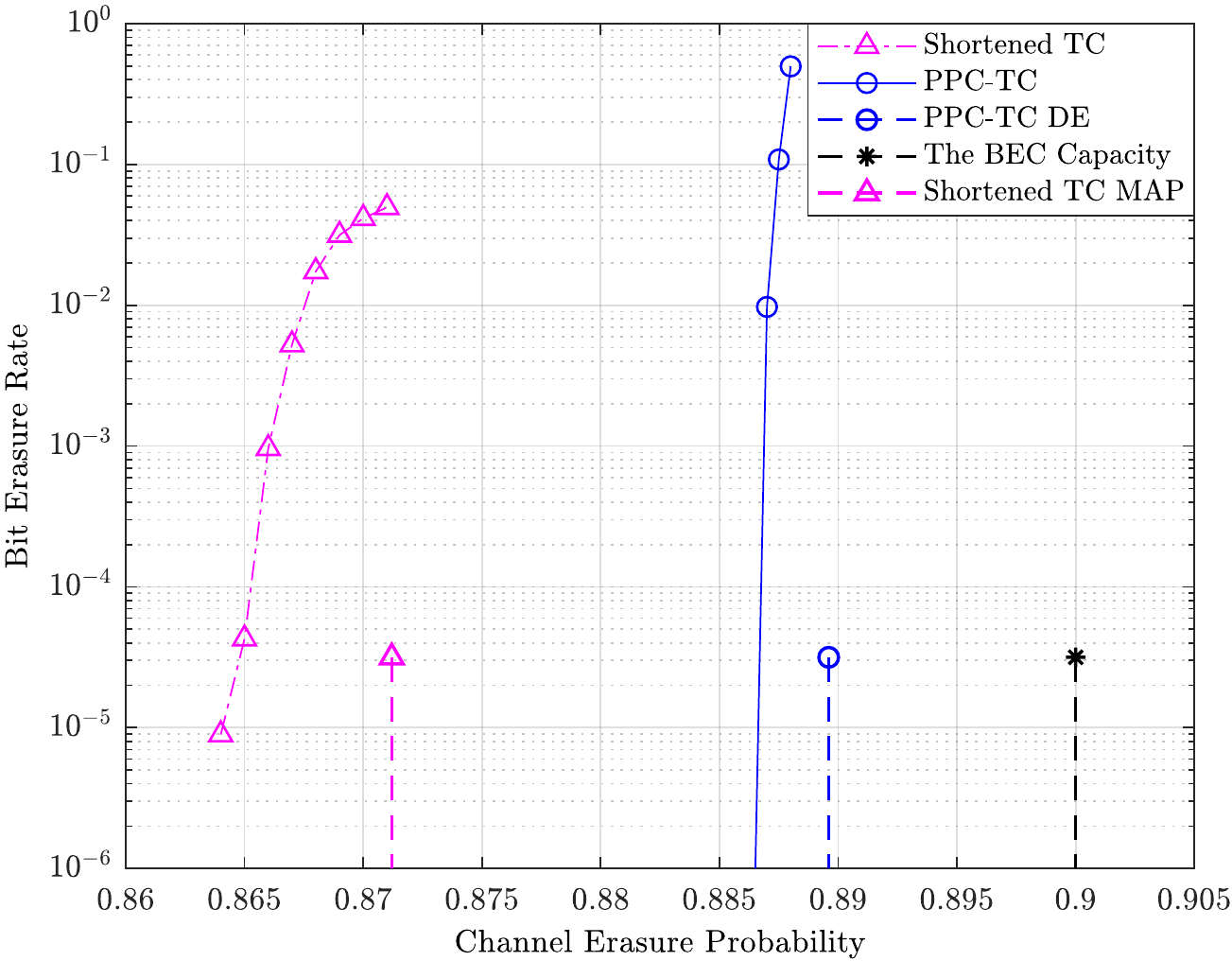}
\caption{Bit erasure rate of PPC-TCs for rate-$1/10$.}
\label{fig:PPC_low}
\end{figure}

To verify the above results, we simulate the error performance for the designed rate-compatible PIC-TCs and PPC-TCs with coupling memory $m=1$ and two code rates $2/3$ and $3/4$. The information length of the component turbo code is $K=100000$ and the coupling length is $L=100$. The parity digits of both codes are randomly punctured. We choose $\lambda=0.5$ for PIC-TCs with rates $2/3$ and $3/4$ and $\lambda=0.2870$ and $\lambda=0.3093$ for PPC-TCs with rates $2/3$ and $3/4$, respectively, according to Table \ref{RC_compare_1}. The error performance of both codes under the FF-FB decoding and the corresponding DE thresholds are shown in Fig. \ref{fig:PIC_punc_rand}. In addition, we also include the MAP threshold for the uncoupled turbo codes (also the iterative decoding threshold of SC-PCCs \cite[Tabel II]{8368318}) as well as the BEC capacity in the figure.

First, it can be observed that all PIC-TCs and PPC-TCs perform close to their respective DE thresholds. In addition, the optimized PPC-TCs have better error performance than the optimized PIC-TCs. By comparing the finite length performance of PIC-TCs and the MAP threshold of uncoupled turbo codes, it is seen that PIC-TCs and PPC-TCs can achieve significant coding gains over SC-PCCs \cite{8368318}.

\subsection{Very Low Rate Performance by Further Increasing the Coupling Ratio}
In this section, we consider the case of very low code rates.

We are particularly interested in the performance of PPC-TCs because very low code rates can be achieved by increasing the coupling ratio while for PIC-TCs lower rate component encoders or shortening is required. It is also worth pointing out that for PPC-TCs $\lambda$ and $\rho$ can be jointly optimized to achieve the best decoding thresholds even in the very low rate regime. Here, we only consider increasing the coupling ratio (without puncturing) for simplicity and we will show that this is sufficient for PPC-TCs to attain close-to-capacity performance at very low code rates. The DE thresholds for the PPC-TCs with rates $1/10, 1/20$ and $1/100$ are reported in Table \ref{table_PPC_low_rate_m}.

\begin{table}[t!]
  \centering
 \caption{DE thresholds for PPC-TCs with $\lambda>1/2$}\label{table_PPC_low_rate_m}
\begin{tabular}{c c c c c c c}
\hline
 Ensemble & Rate  & $\lambda$    & $\epsilon^{(m=1)}_{\text{BP}}$ & $\epsilon^{(m=15)}_{\text{BP}}$ & $\delta^{(m=1)}_{\epsilon}$ &  $\delta^{(m=15)}_{\epsilon}$\\  \hline
 PPC-TC & $1/10$ & $7/9$   & 0.8896 & 0.8990 & 0.0104 & 0.0010 \\

 PPC-TC & $1/20$ & $17/19$   & 0.9411 & 0.9494 & 0.0089 & 0.0006 \\

  PPC-TC &   $1/100$ & $97/99$  & 0.9858 & 0.9894 & 0.0042 & 0.0006 \\
  \hline
\end{tabular}
\end{table}

It is interesting to see that by further increasing the coupling ratio, PPC-TCs can operate even closer to the BEC capacity with a very low code rate compared to the case of smaller coupling ratio shown in Table \ref{table_PIC_m}.

We then show the simulated error performance of the PPC-TC for rate-$1/10$ with $m=1$ and $\lambda = 7/9$ in Fig. \ref{fig:PPC_low}. Again, the error performance and the MAP decoding threshold of a shortened turbo codes with the same code rate are also included.

From Fig. \ref{fig:PPC_low}, similar observation can be made that the actual error performance of PPC-TCs is very close to the DE threshold and significantly better than the uncoupled shortened turbo codes. For spatially coupled turbo-like codes \cite{8368318}, such low rate would require using lower rate component encoders. Hence, PPC-TCs with large coupling ratio $\lambda$ present a simple way to construct very low rate capacity-approaching channel codes.

\subsection{Practical Codeword Length Performance under Sliding Window Decoding}

In this section, we investigate the error performance under sliding window decoding.

\begin{figure}[t!]
	\centering
\includegraphics[width=3.42in,clip,keepaspectratio]{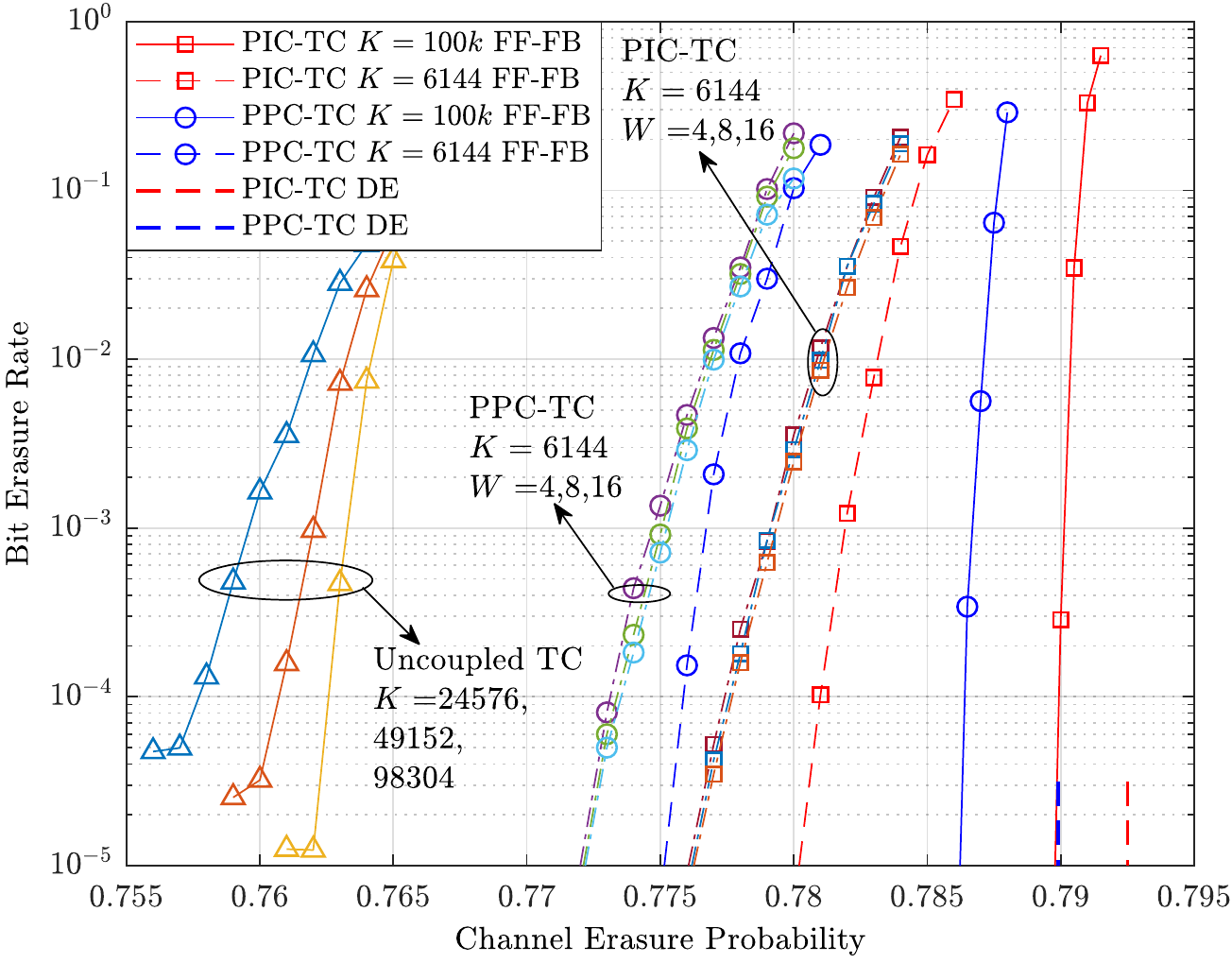}
\caption{Bit erasure rate of PIC-TCs and PPC-TCs for rate-0.2 under sliding window decoding.}
\label{fig:window_decoding_1}
\end{figure}

\begin{figure}[t!]
	\centering
\includegraphics[width=3.42in,clip,keepaspectratio]{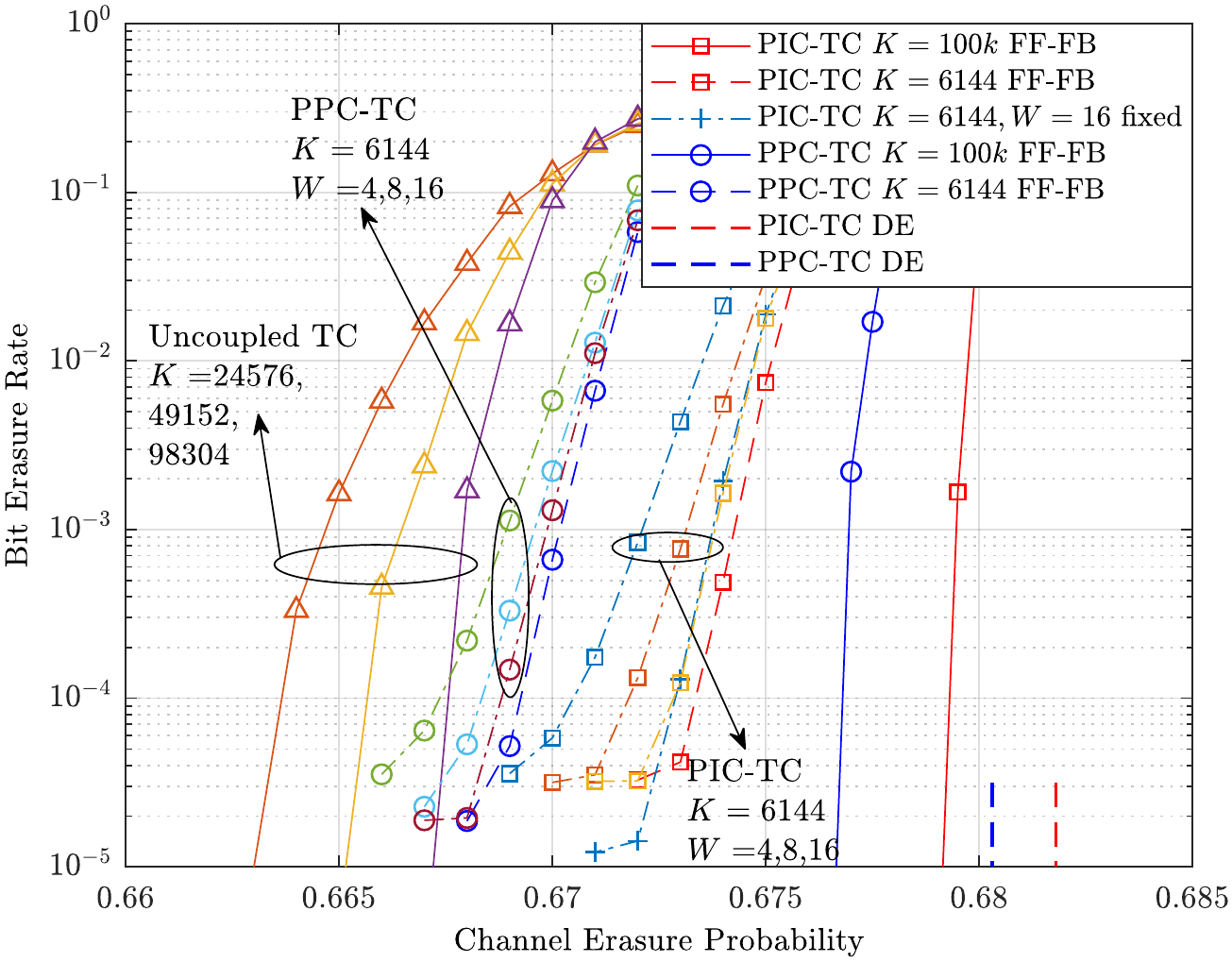}
\caption{Bit erasure rate of PIC-TCs and PPC-TCs for rate-0.3043 under sliding window decoding.}
\label{fig:window_decoding_2}
\end{figure}

We set $K = 6144$, $\lambda = 1/8, 1/2$, and choose the window size $W = 4,8,16$. The corresponding decoding latencies $\mathcal{L} = W \cdot K$ \cite{7296605} are 24576, 49152, 98304 bits. For comparison, we include the error performance of the uncoupled shortened turbo codes with equal rates and decoding latencies as well as the error performance of PIC-TCs and PPC-TCs with $K = 6144$ and $K=100000$ under the FF-FB decoding. The information lengths of the uncoupled turbo codes are 24576, 49152, and 98304, respectively. The BER performance for rates 0.2 and 0.3034 are shown in Figs. \ref{fig:window_decoding_1}-\ref{fig:window_decoding_2}, respectively.


It can be observed that for both PIC-TCs and PPC-TCs, when the coupling ratio is small, the decoding performance under a window size of 8 approaches to that under the FF-FB decoding while for the large coupling ratio increasing the window size from 4 to 16 leads to negligible performance improvement. For both cases, the gap on the decoding performance between the sliding window decoding with $W=4$ and the FF-FB decoding is less than 0.004, which suggests that choosing window size 4 suffices to achieve considerable error performance. Moreover, both PIC-TCs and PPC-TCs show superior waterfall performance over the uncoupled shortened turbo codes with the same or smaller decoding latency. It is also interesting to see that the error floor performance of the codes with larger coupling ratios is better than that with smaller coupling ratios. In light of the decoding performance shown in Sec. \ref{sec:RC_PIC_1}, using of large coupling ratio is more desirable for PIC-TCs and PPC-TCs as the codes with larger $\lambda$ show better waterfall and error floor performance than the codes with smaller $\lambda$.

Since the length of the component codeword is short, the error floor performance is also largely affected by the underlying interleavers. It is possible to design a fixed interleaver to improve the error performance of PIC-TCs and PPC-TCs under short block length. For example, we applied a fixed interleaver from the LTE standard \cite{LTE136212} to the PIC-TC with $\lambda = 1/8$, $W=16$ and show its decoding performance (labeled ``PIC-TC $W = 16$ fixed'') in Fig. \ref{fig:window_decoding_2}. It can be seen that the error floor performance under the LTE interleaver is improved by about half an order of magnitude compared to that under the random interleaver. The design of fixed interleavers and the analysis of the error floor are left for our future work.

\subsection{Performance on the AWGN channel}
In this section, we investigate the performance of PIC-TCs and PPC-TCs on the AWGN channel.

First, we note that directly analyzing the decoding thresholds for the families of spatially coupled turbo-like codes is difficult since the DE equations of the component convolutional codes can no longer be obtained in closed form \cite{8002601}. We also note that Monte Carlo methods can be applied to estimate the decoding thresholds by tracking the densities of a huge number of message samples in each iteration, albeit extremely time-consuming \cite{8437937}. This approach requires significantly large computational effort for spatially coupled codes because their graphs contain different edge types whose message densities have to be considered individually during density evolution. Alternatively, one can estimate the decoding thresholds on the AWGN channel from the thresholds obtained for the BEC channel by using the technique introduced in \cite{chung2000construction,8437937,8631116}. Such an approach is computationally attractive for turbo-like codes since the BEC thresholds can be computed exactly with relatively small effort. Most importantly, it is shown in \cite{8437937,8631116} that the estimated thresholds by this approach are very close to the Monte Carlo thresholds for turbo-like codes.

Let $C_\text{E}(\epsilon) = 1-\epsilon$ be the capacity of the BEC with erasure probability $\epsilon$ and $C_\text{G}(\sigma)$ the capacity of the binary-input AWGN channel whose noise follows $\mathcal{N}(0,\sigma^2)$. Let $\epsilon^*$ and $\sigma^*$ be the decoding thresholds of the code ensembles on the BEC and the AWGN channel, respectively. Based on the observation \cite{chung2000construction,8437937}
\begin{align}
C_\text{G}(\sigma)\approx C_\text{E}(\epsilon),
\end{align}
the following approximation can be obtained
\begin{align}\label{eq:app_awgn_r1a}
\sigma^* \approx C^{-1}_\text{G}(C_\text{E}(\epsilon)) = C^{-1}_\text{G}(1-\epsilon^*).
\end{align}

\begin{figure}[t!]
	\centering
\includegraphics[width=3.42in,clip,keepaspectratio]{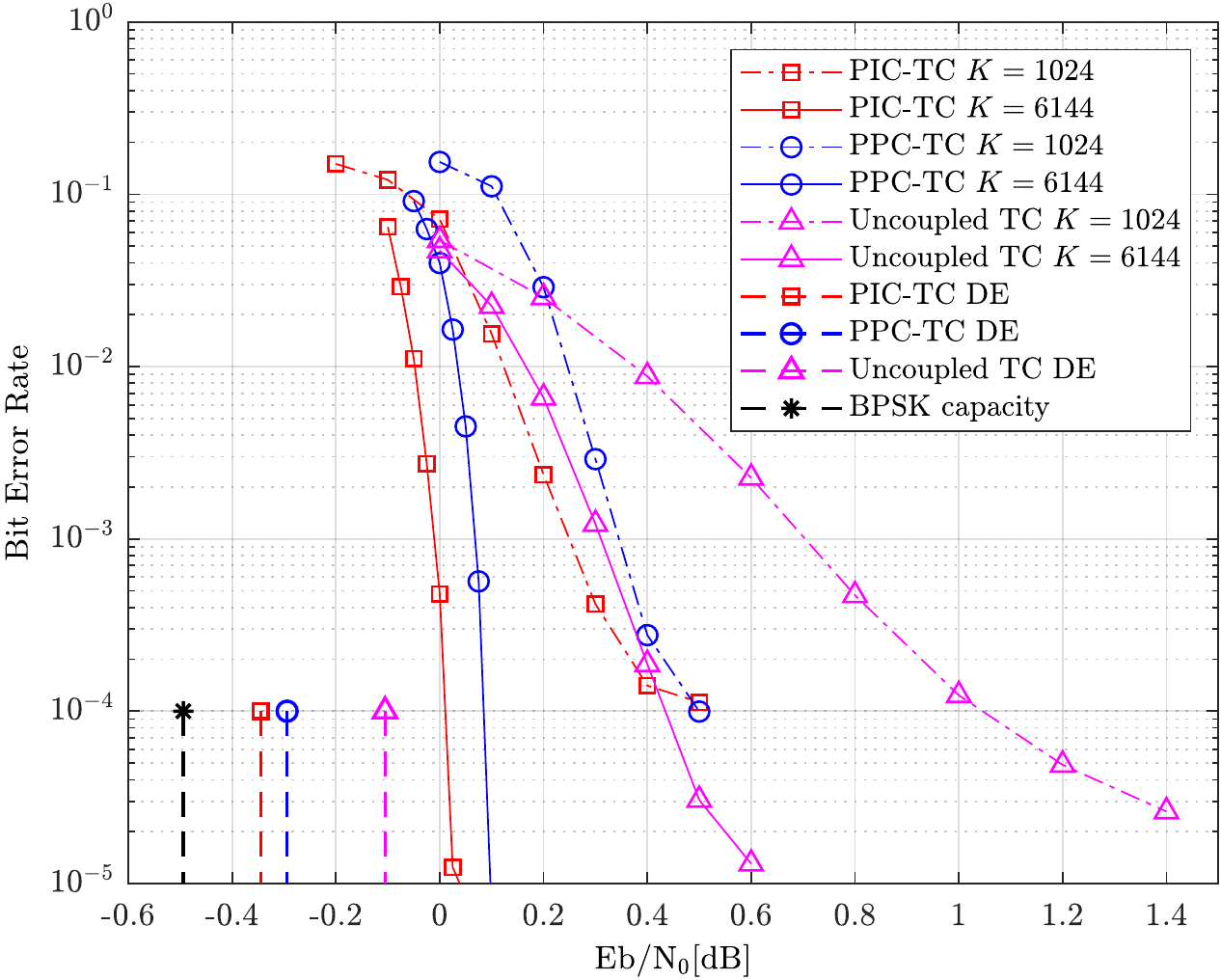}
\caption{Simulation results for rate-1/3 PIC-TCs and PPC-TCs on the AWGN channel.}
\label{fig:AWGN_r1}
\end{figure}

We consider PIC-TCs and PPC-TCs with coupling memory $m=1$ and rate-$1/3$. By using \eqref{eq:app_awgn_r1a}, the AWGN decoding thresholds for PIC-TC and PPC-TC ensembles are 0.345 dB and 0.294 dB, respectively. To simulate the error performance of PIC-TCs and PPC-TCs, we set the coupling length $L=100$ and the information length of the component turbo code $K=1024$ and $K =6144$. We choose $\lambda=0.39$ for PIC-TCs and $\lambda=0.34$ for PPC-TCs according to Table III. The error performance of PIC-TCs, PPC-TCs under FF-FB decoding and uncoupled turbo codes, as well as their corresponding DE thresholds and the capacity of BPSK signaling are shown in Fig. \ref{fig:AWGN_r1}.

It can be seen that at a BER of $10^{-5}$, the PIC-TC and PPC-TC with $K=6144$ operate within 0.4 dB of their respective DE thresholds, which are about 0.1 dB away from the capacity. In addition, the PIC-TC performs slightly better than the PPC-TC by about 0.1 dB in agreement with the results on the BEC (see Table \ref{RC_compare_1}). All the above observations confirm that the predicted decoding thresholds are accurate and the optimal design of $(\lambda,\rho)$ based on the BEC is also effective for the AWGN channel. It is also noteworthy that both codes significantly outperform the uncoupled turbo codes for various component code lengths.

\subsection{Comparisons between PIC-TCs and PPC-TCs}

Finally, we give a detailed comparison between PIC-TCs and PPC-TCs. The main difference is that the coupled information sequence and the parity sequence are protected by two component turbo codewords in PIC-TCs and PPC-TCs, respectively. As a result, PIC-TCs can enjoy parallel encoding by encoding all the component turbo codeword at the same time while PPC-TCs have to encode each component codeword one by one. That being said, the design of PPC-TCs is more flexible as it can be easily extended to very low rates by increasing the coupling ratio. The extension for PIC-TCs to very low rates is nontrivial since the coupling ratio cannot be larger than 0.5. Most importantly, PPC-TCs have better decoding thresholds than PIC-TCs when the code rate is greater than $1/3$ and the coupling memory is large. It can be seen from Table III and Table IV that PPC-TCs with a joint design of coupling ratio and puncturing ratio approach the BEC capacity for a wide range of code rates and their performance gap to the capacity almost vanishes when the coupling memory is large.

\section{Concluding Remarks}\label{sec:conclude}
We investigated the impact of partial coupling on the iterative decoding thresholds of turbo codes. We first studied the performance of PIC-TCs under the BEC and also proposed PPC-TCs. We provided the encoding and decoding methods for PIC-TCs and PPC-TCs with coupling memory $m \geq 1$. We derived the exact DE equations for any given coupling memory and coupling ratio. Rate-compatible designs for PIC-TCs and PPC-TCs based on jointly designing the puncturing ratio and the coupling ratio are also proposed. Both our theoretical analysis and simulation results show that existing uncoupled turbo codes can be significantly benefit from partially information and partially parity coupling. In particular, the proposed PPC-TCs have better decoding thresholds than existing spatially coupled turbo-like codes and can operate within 0.0002 to the BEC capacity for a wide range of code rates without changing the encoding and decoding architecture of the component turbo codes.

\appendices
\section{Density Evolution for PIC-TCs with $m\geq1$}\label{app:PIC}

For $m\geq 1$, the erasure probability of $\mathbf{u}_t$ depends on the erasure probability of the coupled information $\mathbf{u}_{t-m,t},\ldots,\mathbf{u}_{t-1,t}$ and $\mathbf{u}_{t,t+1},\ldots,\mathbf{u}_{t,t+m}$. Hence, the average erasure probability of $\mathbf{u}_t$ to $f^{\text{U}}$ at time $t$ from \eqref{eq:1} needs to be modified to
\begin{align}
\bar{p}^{(i)}_{\text{L},t}
=& \epsilon\Bigg(\frac{\lambda}{m}\sum\limits_{j=1}^m p^{(i)}_{\text{L},t-j}\cdot p^{(i-1)}_{\text{U},t-j}\cdot p^{(i)}_{\text{L},t}+ (1-2\lambda) p^{(i)}_{\text{L},t} \nonumber \\
 &+ \frac{\lambda}{m}\sum\limits_{j=1}^m p^{(i)}_{\text{L},t}\cdot p^{(i-1)}_{\text{U},t+j}\cdot p^{(i)}_{\text{L},t+j}\Bigg) \nonumber \\
=& \epsilon\cdot p^{(i)}_{\text{L},t}\Bigg(\frac{\lambda}{m}\sum\limits_{j=1}^m \left(p^{(i-1)}_{\text{U},t-j}\cdot p^{(i)}_{\text{L},t-j}+ p^{(i-1)}_{\text{U},t+j}\cdot p^{(i)}_{\text{L},t+j}\right) \nonumber \\
&+ (1-2\lambda)\Bigg),
\end{align}
where $\frac{\lambda}{m}$ is the portion of the coupled information sequence shared between two CBs at different time instances.

Similarly, the average erasure probability from $\mathbf{u}_t$ to node $f^{\text{L}}$ at time $t$ from \eqref{eq:2} is modified to
\begin{align}\label{eq:mPIC-TC}
\bar{p}^{(i)}_{\text{U},t} =& \epsilon\Bigg(\frac{\lambda}{m}\sum\limits_{j=1}^m p^{(i-1)}_{\text{L},t-j}\cdot p^{(i)}_{\text{U},t-j}\cdot p^{(i)}_{\text{U},t}+ (1-2\lambda) p^{(i)}_{\text{U},t}   \nonumber \\
&+ \frac{\lambda}{m}\sum\limits_{j=1}^m p^{(i)}_{\text{U},t}\cdot p^{(i-1)}_{\text{L},t+j}\cdot p^{(i)}_{\text{U},t+j}\Bigg) \nonumber \\\
=& \epsilon\cdot p^{(i)}_{\text{U},t}\Bigg(\frac{\lambda}{m}\sum\limits_{j=1}^m \left(p^{(i)}_{\text{U},t-j}\cdot p^{(i-1)}_{\text{L},t-j} + p^{(i)}_{\text{U},t+j}\cdot p^{(i-1)}_{\text{L},t+j}\right) \nonumber \\
&+ (1-2\lambda)\Bigg).
\end{align}
The rest of the DE equations remain the same as those for the case of $m=1$ in Section \ref{sec:DE_PIC}.

\section{Density Evolution for PPC-TCs with $m\geq1$}\label{app:PPC}

For $m\geq 1$, the average erasure probability of $\mathbf{u}_t$ to $f^{\text{U}}$ at time $t$ is the weighted sum of the erasure probability from nodes $\mathbf{v}^{\text{U}}_{t-j},\mathbf{v}^{\text{L}}_{t-j},\mathbf{u}'_t$ to $f^{\text{U}}$ for $j \in \{1,\ldots,m\}$. Hence, \eqref{eq:PPC_DE_1} is modified to
\begin{align}\label{eq:PPC_DE_m_1}
\bar{p}^{(i)}_{\text{L},t}
 =&\epsilon\Bigg(\frac{\lambda_q^{\text{U}}}{m}\sum\limits_{j=1}^m q^{(i-1)}_{\text{U},t-j}\cdot  p^{(i)}_{\text{L},t} +  (1 - \lambda_q^{\text{U}}-\lambda_q^{\text{L}}) p^{(i)}_{\text{L},t} \nonumber \\
 &+ \frac{\lambda_q^{\text{L}}}{m}\sum\limits_{j=1}^m q^{(i)}_{\text{L},t-j}\cdot  p^{(i)}_{\text{L},t}\Bigg) \nonumber \\
 =&  \epsilon \cdot p^{(i)}_{\text{L},t}\Bigg(\sum\limits_{j=1}^m \left(\frac{\lambda_q^{\text{U}}}{m} \cdot q^{(i-1)}_{\text{U},t-j}  + \frac{\lambda_q^{\text{L}}}{m} \cdot q^{(i)}_{\text{L},t-j}\right)\nonumber \\
 &+ 1 - \lambda_q^{\text{U}}-\lambda_q^{\text{L}} \Bigg),
\end{align}
where $\frac{\lambda_q^{\text{U}}}{m}$ and $\frac{\lambda_q^{\text{L}}}{m}$ are the portions of the coupled upper and lower parity sequence, respectively, shared between two CBs at different time instances. In addition, the average erasure probability of $\mathbf{v}^{\text{U}}_{t}$ to $f^{\text{U}}$ from \eqref{eq:PPC_DE_1a} is modified to
\begin{align}\label{eq:PPC_DE_m_1a}
\bar{q}^{(i)}_{\text{U},t} = \epsilon\left(\left(1-\lambda_q^{\text{U}}\right)+\frac{\lambda_q^{\text{U}}}{m}\sum\limits_{j=1}^m( p^{(i)}_{\text{L},t+j}\cdot p^{(i-1)}_{\text{U},t+j})\right).
\end{align}

Likewise, the average erasure probability from $\mathbf{u}_t$ to node $f^{\text{L}}$ at time $t$ from \eqref{eq:PPC_DE_2} is modified to
\begin{align}\label{eq:PPC_DE_m_2}
\bar{p}^{(i)}_{\text{U},t} =& \epsilon\Bigg(\frac{\lambda_p^{\text{U}}}{m}\sum\limits_{j=1}^m q^{(i)}_{\text{U},t-j}\cdot p^{(i)}_{\text{U},t} +  (1 -  \lambda_p^{\text{U}}-\lambda_p^{\text{L}}) p^{(i)}_{\text{U},t} \nonumber \\
&+ \frac{\lambda_p^{\text{L}}}{m} \sum\limits_{j=1}^m p^{(i)}_{\text{U},t}\cdot q^{(i-1)}_{\text{L},t-j}\Bigg) \nonumber \\
 =& \epsilon \cdot p^{(i)}_{\text{U},t}\Bigg(\sum\limits_{j=1}^m \left(\frac{\lambda_p^{\text{U}}}{m}\cdot q^{(i)}_{\text{U},t-1} +\frac{\lambda_p^{\text{L}}}{m}\cdot q^{(i-1)}_{\text{L},t-1}\right) \nonumber \\
 &+  1 -  \lambda_p^{\text{U}}-\lambda_p^{\text{L}} \Bigg),
\end{align}
and the average erasure probability of $\mathbf{v}^{\text{L}}_{t}$ to $f^{\text{L}}$ from \eqref{eq:PPC_DE_2a} is
\begin{align}\label{eq:PPC_DE_m_2a}
\bar{q}^{(i)}_{\text{L},t} = \epsilon\left((1-\lambda_q^{\text{L}})+\frac{\lambda_q^{\text{L}}}{m}\sum\limits_{j=1}^m( p^{(i)}_{\text{U},t+j}\cdot p^{(i-1)}_{\text{L},t+j})\right).
\end{align}

The rest of the DE equations remain the same as those for the case of $m=1$ in Section \ref{sec:DE_PPC}.

\bibliographystyle{IEEEtran}
\bibliography{MinQiu}

\begin{thebibliography}{10}
\providecommand{\url}[1]{#1}
\csname url@samestyle\endcsname
\providecommand{\newblock}{\relax}
\providecommand{\bibinfo}[2]{#2}
\providecommand{\BIBentrySTDinterwordspacing}{\spaceskip=0pt\relax}
\providecommand{\BIBentryALTinterwordstretchfactor}{4}
\providecommand{\BIBentryALTinterwordspacing}{\spaceskip=\fontdimen2\font plus
\BIBentryALTinterwordstretchfactor\fontdimen3\font minus
  \fontdimen4\font\relax}
\providecommand{\BIBforeignlanguage}[2]{{%
\expandafter\ifx\csname l@#1\endcsname\relax
\typeout{** WARNING: IEEEtran.bst: No hyphenation pattern has been}%
\typeout{** loaded for the language `#1'. Using the pattern for}%
\typeout{** the default language instead.}%
\else
\language=\csname l@#1\endcsname
\fi
#2}}
\providecommand{\BIBdecl}{\relax}
\BIBdecl

\bibitem{8989359}
M.~Qiu, X.~Wu, and J.~Yuan, ``Density evolution analysis of partially
  information coupled turbo codes on the erasure channel,'' in \emph{Inf.
  Theory Workshop (ITW)}, Aug. 2019, pp. 1--5.

\bibitem{LDPCCthesis}
A.~{J. Felstr\"om}, \emph{Soft Iterative Decoding of Low-Density Convolutional
  Codes}.\hskip 1em plus 0.5em minus 0.4em\relax Engineering Licenciate Thesis,
  Lund University, Oct. 1997.

\bibitem{782171}
A.~{J. Felstr\"om} and K.~S. {Zigangirov}, ``Time-varying periodic
  convolutional codes with low-density parity-check matrix,'' \emph{IEEE Trans.
  Inf. Theory}, vol.~45, no.~6, pp. 2181--2191, Sep. 1999.

\bibitem{7340116}
A.~{Graell i Amat}, C.~{H\"{a}ger}, F.~{Br\"{a}nnstr\"{o}m}, and E.~{Agrell},
  ``Spatially-coupled codes for optical communications: state-of-the-art and
  open problems,'' in \emph{Opto-Electronics Commun. Conf.}, 2015, pp. 1--3.

\bibitem{7553579}
Y.~Xie, L.~Yang, P.~Kang, and J.~Yuan, ``Euclidean geometry-based spatially
  coupled {LDPC} codes for storage,'' \emph{IEEE J. Sel. Areas Commun.},
  vol.~34, no.~9, pp. 2498--2509, Sep. 2016.

\bibitem{1056404}
R.~Tanner, ``A recursive approach to low complexity codes,'' \emph{IEEE Trans.
  Inf. Theory}, vol.~27, no.~5, pp. 533--547, Sep. 1981.

\bibitem{Gallager63low-densityparity-check}
R.~G. Gallager, ``Low-density parity-check codes,'' \emph{MIT Press}, 1963.

\bibitem{5571910}
M.~Lentmaier, A.~Sridharan, D.~J. Costello, and K.~S. Zigangirov, ``Iterative
  decoding threshold analysis for {LDPC} convolutional codes,'' \emph{IEEE
  Trans. Inf. Theory}, vol.~56, no.~10, pp. 5274--5289, Oct. 2010.

\bibitem{5695130}
S.~Kudekar, T.~J. Richardson, and R.~L. Urbanke, ``Threshold saturation via
  spatial coupling: Why convolutional {LDPC} ensembles perform so well over the
  {BEC},'' \emph{IEEE Trans. Inf. Theory}, vol.~57, no.~2, pp. 803--834, Feb.
  2011.

\bibitem{7130575}
Y.~{Liu}, Y.~{Li}, and Y.~{Chi}, ``Spatially coupled {LDPC} codes constructed
  by parallelly connecting multiple chains,'' \emph{IEEE Commun. Lett.},
  vol.~19, no.~9, pp. 1472--1475, 2015.

\bibitem{7152893}
D.~G.~M. {Mitchell}, M.~{Lentmaier}, and D.~J. {Costello}, ``Spatially coupled
  {LDPC} codes constructed from protographs,'' \emph{IEEE Trans. Inf. Theory},
  vol.~61, no.~9, pp. 4866--4889, Sep. 2015.

\bibitem{7533500}
Y.~{Chi}, Y.~{Li}, G.~{Song}, and Y.~{Sun}, ``Partially repeated {SC-LDPC}
  codes for multiple-access channel,'' \emph{IEEE Commun. Lett.}, vol.~20,
  no.~10, pp. 1947--1950, 2016.

\bibitem{6325197}
A.~{Yedla}, Y.~{Jian}, P.~S. {Nguyen}, and H.~D. {Pfister}, ``A simple proof of
  threshold saturation for coupled scalar recursions,'' in \emph{Proc. Int.
  Symp. Turbo Codes Iterative Inf. Process (ISTC)}, 2012, pp. 51--55.

\bibitem{6589171}
S.~{Kudekar}, T.~{Richardson}, and R.~L. {Urbanke}, ``Spatially coupled
  ensembles universally achieve capacity under belief propagation,'' \emph{IEEE
  Trans. Inf. Theory}, vol.~59, no.~12, pp. 7761--7813, Dec. 2013.

\bibitem{6912949}
S.~{Kumar}, A.~J. {Young}, N.~{Macris}, and H.~D. {Pfister}, ``Threshold
  saturation for spatially coupled {LDPC} and {LDGM} codes on {BMS} channels,''
  \emph{IEEE Trans. Inf. Theory}, vol.~60, no.~12, pp. 7389--7415, 2014.

\bibitem{6374679}
A.~R. {Iyengar}, P.~H. {Siegel}, R.~L. {Urbanke}, and J.~K. {Wolf}, ``Windowed
  decoding of spatially coupled codes,'' \emph{IEEE Trans. Inf. Theory},
  vol.~59, no.~4, pp. 2277--2292, 2013.

\bibitem{7265214}
E.~Ankan, N.~ul~Hassan, M.~Lentmaier, G.~Montorsi, and J.~Sayir, ``Challenges
  and some new directions in channel coding,'' \emph{J. of Commun. and
  Networks}, vol.~17, no.~4, pp. 328--338, Aug. 2015.

\bibitem{Smith12}
B.~P. Smith, A.~Farhood, A.~Hunt, F.~R. Kschischang, and J.~Lodge, ``Staircase
  codes: {FEC} for 100 {G}b/s {OTN},'' \emph{J. Lightw. Technol.}, vol.~30,
  no.~1, pp. 110--117, Jan. 2012.

\bibitem{8425763}
M.~Qiu, L.~Yang, Y.~Xie, and J.~Yuan, ``Terminated staircase codes for {NAND}
  flash memories,'' \emph{IEEE Trans. Commun.}, vol.~66, no.~12, pp.
  5861--5875, Dec. 2018.

\bibitem{8002601}
S.~Moloudi, M.~Lentmaier, and A.~{Graell i Amat}, ``Spatially coupled
  turbo-like codes,'' \emph{IEEE Trans. Inf. Theory}, vol.~63, no.~10, pp.
  6199--6215, Oct. 2017.

\bibitem{8631116}
S.~{Moloudi}, M.~{Lentmaier}, and A.~{Graell i Amat}, ``Spatially coupled
  turbo-like codes: A new trade-off between waterfall and error floor,''
  \emph{IEEE Trans. Commun.}, vol.~67, no.~5, pp. 3114--3123, 2019.

\bibitem{910572}
F.~R. Kschischang, B.~J. Frey, and H.~A. Loeliger, ``Factor graphs and the
  sum-product algorithm,'' \emph{IEEE Trans. Inf. Theory}, vol.~47, no.~2, pp.
  498--519, Feb. 2001.

\bibitem{397441}
C.~Berrou, A.~Glavieux, and P.~Thitimajshima, ``Near shannon limit
  error-correcting coding and decoding: Turbo-codes. 1,'' in \emph{Proc. IEEE
  Int. Conf. Commun. (ICC)}, vol.~2, May 1993, pp. 1064--1070.

\bibitem{Vucetic:2000:TCP:352869}
B.~Vucetic and J.~Yuan, \emph{Turbo Codes: Principles and Applications}.\hskip
  1em plus 0.5em minus 0.4em\relax Norwell, MA, USA: Kluwer Academic
  Publishers, 2000.

\bibitem{669119}
S.~{Benedetto}, D.~{Divsalar}, G.~{Montorsi}, and F.~{Pollara}, ``Serial
  concatenation of interleaved codes: performance analysis, design, and
  iterative decoding,'' \emph{IEEE Trans. Inf. Theory}, vol.~44, no.~3, pp.
  909--926, May 1998.

\bibitem{5361461}
W.~{Zhang}, M.~{Lentmaier}, K.~S. {Zigangirov}, and D.~J. {Costello}, ``Braided
  convolutional codes: A new class of turbo-like codes,'' \emph{IEEE Trans.
  Inf. Theory}, vol.~56, no.~1, pp. 316--331, Jan. 2010.

\bibitem{8368318}
L.~Yang, Y.~Xie, X.~Wu, J.~Yuan, X.~Cheng, and L.~Wan, ``Partially
  information-coupled turbo codes for {LTE} systems,'' \emph{IEEE Trans.
  Commun.}, vol.~66, no.~10, pp. 4381--4392, Oct. 2018.

\bibitem{4907407}
A.~Larmo, M.~Lindström, M.~Meyer, G.~Pelletier, J.~Torsner, and H.~Wiemann,
  ``The {LTE} link-layer design,'' \emph{IEEE commun. Mag.}, vol.~47, no.~4,
  pp. 52--59, Apr. 2009.

\bibitem{6824752}
J.~G. Andrews, S.~Buzzi, W.~Choi, S.~V. Hanly, A.~Lozano, A.~C.~K. Soong, and
  J.~C. Zhang, ``What will 5{G} be?'' \emph{IEEE J. Sel. Areas Commun.},
  vol.~32, no.~6, pp. 1065--1082, Jun. 2014.

\bibitem{8301547}
L.~{Yang}, Y.~{Xie}, J.~{Yuan}, X.~{Cheng}, and L.~{Wan}, ``Chained {LDPC}
  codes for future communication systems,'' \emph{IEEE Commun. Lett.}, vol.~22,
  no.~5, pp. 898--901, 2018.

\bibitem{8470926}
X.~{Wu}, L.~{Yang}, Y.~{Xie}, and J.~{Yuan}, ``Partially information coupled
  polar codes,'' \emph{IEEE Access}, vol.~6, pp. 63\,689--63\,702, 2018.

\bibitem{957394}
S.~ten Brink, ``Convergence behavior of iteratively decoded parallel
  concatenated codes,'' \emph{IEEE Trans. Commun.}, vol.~49, no.~10, pp.
  1727--1737, Oct. 2001.

\bibitem{910578}
T.~J. Richardson, M.~A. Shokrollahi, and R.~L. Urbanke, ``Design of
  capacity-approaching irregular low-density parity-check codes,'' \emph{IEEE
  Trans. Inf. Theory}, vol.~47, no.~2, pp. 619--637, Feb. 2001.

\bibitem{1055186}
L.~Bahl, J.~Cocke, F.~Jelinek, and J.~Raviv, ``Optimal decoding of linear codes
  for minimizing symbol error rate,'' \emph{IEEE Trans. Inf. Theory}, vol.~20,
  no.~2, pp. 284--287, Mar. 1974.

\bibitem{370145}
M.~R. {Best}, M.~V. {Burnashev}, Y.~{Levy}, A.~{Rabinovich}, P.~C. {Fishburn},
  A.~R. {Calderbank}, and D.~J. {Costello}, ``On a technique to calculate the
  exact performance of a convolutional code,'' \emph{IEEE Trans. Inf. Theory},
  vol.~41, no.~2, pp. 441--447, 1995.

\bibitem{1258535}
B.~M. {Kurkoski}, P.~H. {Siegel}, and J.~K. {Wolf}, ``Exact probability of
  erasure and a decoding algorithm for convolutional codes on the binary
  erasure channel,'' in \emph{Proc. IEEE Globecom}, vol.~3, Dec. 2003, pp.
  1741--1745.

\bibitem{1523540}
C.~{Measson}, R.~{Urbanke}, A.~{Montanari}, and T.~{Richardson}, ``Maximum a
  posteriori decoding and turbo codes for general memoryless channels,'' in
  \emph{Proc. IEEE Int. Symp. Inf. Theory (ISIT)}, Sep. 2005, pp. 1241--1245.

\bibitem{7353121}
D.~G.~M. {Mitchell}, M.~{Lentmaier}, A.~E. {Pusane}, and D.~J. {Costello},
  ``Randomly punctured {LDPC} codes,'' \emph{IEEE J. Sel. Areas Commun.},
  vol.~34, no.~2, pp. 408--421, 2016.

\bibitem{7296605}
C.~{Rachinger}, J.~B. {Huber}, and R.~R. {M\"{u}ller}, ``Comparison of
  convolutional and block codes for low structural delay,'' \emph{IEEE Trans.
  Commun.}, vol.~63, no.~12, pp. 4629--4638, 2015.

\bibitem{LTE136212}
ETSI, ``{LTE};. evolved universal terrestrial radio access ({E-UTRA});.
  multiplexing and channel coding,'' {European Telecommunications Standards
  Institute (ETSI)}, TR {ETSI TS 136 212 V14.2.0}, Apr. 2017.

\bibitem{8437937}
M.~U. {Farooq}, S.~{Moloudi}, and M.~{Lentmaier}, ``Thresholds of braided
  convolutional codes on the {AWGN} channel,'' in \emph{Proc. IEEE Int. Symp.
  Inf. Theory (ISIT)}, 2018, pp. 1375--1379.

\bibitem{chung2000construction}
S.-Y. Chung, ``On the construction of some capacity-approaching coding
  schemes,'' Ph.D. dissertation, Massachusetts Institute of Technology, 2000.

\end{thebibliography}

\end{document}